\PassOptionsToPackage{unicode}{hyperref}
\PassOptionsToPackage{hyphens}{url}
\PassOptionsToPackage{dvipsnames,svgnames,x11names}{xcolor}
\documentclass[
  american,
]{article}
\usepackage{xcolor}
\usepackage[top=10pc,bottom=10pc,left=11pc,right=11pc,heightrounded]{geometry}
\usepackage{amsmath,amssymb}
\usepackage{iftex}
\ifPDFTeX
  \usepackage[T1]{fontenc}
  \usepackage[utf8]{inputenc}
  \usepackage{textcomp} 
\else 
  \usepackage{unicode-math} 
  \defaultfontfeatures{Scale=MatchLowercase}
  \defaultfontfeatures[\rmfamily]{Ligatures=TeX,Scale=1}
\fi
\usepackage{lmodern}
\ifPDFTeX\else
\fi
\IfFileExists{upquote.sty}{\usepackage{upquote}}{}
\IfFileExists{microtype.sty}{
  \usepackage[]{microtype}
  \UseMicrotypeSet[protrusion]{basicmath} 
}{}

\usepackage{longtable,booktabs,array}
\usepackage{calc} 
\usepackage{etoolbox}
\makeatletter
\patchcmd\longtable{\par}{\if@noskipsec\mbox{}\fi\par}{}{}
\makeatother
\IfFileExists{footnotehyper.sty}{\usepackage{footnotehyper}}{\usepackage{footnote}}
\makesavenoteenv{longtable}
\usepackage{graphicx}
\makeatletter
\newsavebox\pandoc@box
\newcommand*\pandocbounded[1]{
  \sbox\pandoc@box{#1}%
  \Gscale@div\@tempa{\textheight}{\dimexpr\ht\pandoc@box+\dp\pandoc@box\relax}%
  \Gscale@div\@tempb{\linewidth}{\wd\pandoc@box}%
  \ifdim\@tempb\p@<\@tempa\p@\let\@tempa\@tempb\fi
  \ifdim\@tempa\p@<\p@\scalebox{\@tempa}{\usebox\pandoc@box}%
  \else\usebox{\pandoc@box}%
  \fi%
}
\def\fps@figure{htbp}
\makeatother

\NewDocumentCommand\citeproctext{}{}
\NewDocumentCommand\citeproc{mm}{%
  \begingroup\def\citeproctext{#2}\cite{#1}\endgroup}
\makeatletter
 \let\@cite@ofmt\@firstofone
 \def\@biblabel#1{}
 \def\@cite#1#2{{#1\if@tempswa , #2\fi}}
\makeatother
\newlength{\cslhangindent}
\newlength{\csllabelwidth}
\newenvironment{CSLReferences}[2] 
 {\begin{list}{}{%
  \setlength{\itemindent}{0pt}
  \setlength{\leftmargin}{0pt}
  \setlength{\parsep}{0pt}
  \ifodd #1
   \setlength{\leftmargin}{\cslhangindent}
   \setlength{\itemindent}{-1\cslhangindent}
  \fi
  \setlength{\itemsep}{#2\baselineskip}}}
 {\end{list}}
\usepackage{calc}

\newcommand{\CSLLeftMargin}[1]{\parbox[t]{\csllabelwidth}{\strut#1\strut}}
\newcommand{\CSLRightInline}[1]{\parbox[t]{\linewidth - \csllabelwidth}{\strut#1\strut}}

\ifLuaTeX
\usepackage[bidi=basic,shorthands=off]{babel}
\else
\usepackage[bidi=default,shorthands=off]{babel}
\fi
\ifLuaTeX
  \usepackage{selnolig} 
\fi

\providecommand{\tightlist}{%
  \setlength{\itemsep}{0pt}\setlength{\parskip}{0pt}}

\usepackage[all,defaultlines=2]{nowidow}

\usepackage{enumitem}
\setlist[1]{labelindent=\parindent}
\setlist[itemize]{leftmargin=*}
\setlist[enumerate]{leftmargin=*}

\setlist[description]{style=unboxed}

\usepackage[titles]{tocloft}

\AtBeginEnvironment{longtable}{\setlist[itemize]{nosep, wide=0pt, leftmargin=*, before=\vspace*{-\baselineskip}, after=\vspace*{-\baselineskip}}}

\usepackage{setspace}

\usepackage[overload]{textcase}
\usepackage[runin]{abstract}

\abslabeldelim{\quad}
\usepackage{titling}
\pretitle{\par\vskip 5em \begin{flushleft}\LARGE\sffamily\bfseries}
\posttitle{\par\end{flushleft}\vskip 0.75em}

\renewcommand{\and}{\end{tabular} \hskip 3em \begin{tabular}[t]{@{\hspace{0em}}l@{}}}
\preauthor{\begin{flushleft}
           \lineskip 1.5em 
           \begin{tabular}[t]{@{\hspace{0em}}l@{}}}
\postauthor{\end{tabular}\par\end{flushleft}}

\predate{}
\postdate{}

\newcommand{\published}[1]{%
   \gdef\puB{#1}}
   \newcommand{\puB}{}

\usepackage{titlesec}
\titleformat*{\section}{\Large\sffamily\bfseries\raggedright}
\titleformat*{\subsection}{\large\sffamily\bfseries\raggedright}
\titleformat*{\subsubsection}{\normalsize\sffamily\bfseries\raggedright}
\titleformat*{\paragraph}{\small\sffamily\bfseries\raggedright}

\titlespacing*{\section}{0em}{2em}{0.1em}
\titlespacing*{\subsection}{0em}{1.25em}{0.1em}
\titlespacing*{\subsubsection}{0em}{0.75em}{0em}

\usepackage[bottom, flushmargin]{footmisc}

\usepackage[font={small,sf}, labelfont={small,sf,bf}]{caption}

\usepackage{pdflscape}
\newcommand{\blandscape}{\begin{landscape}}
\newcommand{\elandscape}{\end{landscape}}

\makeatletter
\@ifpackageloaded{caption}{}{\usepackage{caption}}
\AtBeginDocument{%
\ifdefined\contentsname
  \renewcommand*\contentsname{Table of contents}
\else
  \newcommand\contentsname{Table of contents}
\fi
\ifdefined\listfigurename
  \renewcommand*\listfigurename{List of Figures}
\else
  \newcommand\listfigurename{List of Figures}
\fi
\ifdefined\listtablename
  \renewcommand*\listtablename{List of Tables}
\else
  \newcommand\listtablename{List of Tables}
\fi
\ifdefined\figurename
  \renewcommand*\figurename{Figure}
\else
  \newcommand\figurename{Figure}
\fi
\ifdefined\tablename
  \renewcommand*\tablename{Table}
\else
  \newcommand\tablename{Table}
\fi
}
\@ifpackageloaded{float}{}{\usepackage{float}}
\floatstyle{ruled}
\@ifundefined{c@chapter}{\newfloat{codelisting}{h}{lop}}{\newfloat{codelisting}{h}{lop}[chapter]}
\floatname{codelisting}{Listing}

\makeatother
\makeatletter
\@ifpackageloaded{caption}{}{\usepackage{caption}}
\@ifpackageloaded{subcaption}{}{\usepackage{subcaption}}
\makeatother
\usepackage{bookmark}
\IfFileExists{xurl.sty}{\usepackage{xurl}}{} 
\hypersetup{
  pdftitle={OpenID for European Digital Identity},
  pdfauthor={Wouter Termont; Beatriz Esteves},
  pdflang={en-US},
  colorlinks=true,
  linkcolor={DarkSlateBlue},
  filecolor={Maroon},
  citecolor={DarkSlateBlue},
  urlcolor={DarkSlateBlue},
  pdfcreator={LaTeX via pandoc}}

\usepackage{orcidlink}  

\title{OpenID for European Digital Identity\thanks{Research funded by
\href{https://solidlab.be/}{SolidLab Vlaanderen} (Flemish Gov., EWI \&
RRF project VV023/10)}}

\usepackage{etoolbox}
\makeatletter
\providecommand{\subtitle}[1]{
  \apptocmd{\@title}{\par {\vskip 0.25em \large #1 \par}}{}{}
}
\makeatother
\subtitle{An architectural analysis of user-centric identity management}

\author{
{\large Wouter Termont~\orcidlink{0000-0002-2968-1394}}%
\thanks{Corresponding author.} \\%
Ghent University -- imec \\%
{\footnotesize \url{wouter.termont@ugent.be}} \and
{\large Beatriz Esteves~\orcidlink{0000-0003-0259-7560}}%
 \\%
Ghent University -- imec \\%
{\footnotesize \url{beatriz.esteves@ugent.be}} \and
}

\date{}

\usepackage{url}
\begin{document}
\published{\textbf{Monday, January 19, 2026}}

\maketitle

\begin{abstract}
Recent European efforts around digital identity -- the EUDI regulation
and its OpenID architecture -- aim high to provide an EU-wide
authentication framework. However, its current technical and legislative
architecture are based on a limited conceptualization of identity. None
of the legal and technical texts involved explicitly define this central
term; and their implicit model of the concept does not go beyond a
digitalization of identity cards and similar documents. Based on several
other standards, we therefore propose a deeper, explicit definition.
Grounded in this definition, we identify several issues in the design of
OpenID4VCI and OpenID4VP: (1) available credentials are limited to
static, preset configurations, flexible in packaging format but not in
semantic content; (2) querying claims in credentials happens through a
custom, format-specific language; and (3) credentials can only be about
a (single) principal user of a service, who must (synchronously)
interact for each transaction. These problems limit the kind of
credentials that can be issued, the way specific information can be
requested, and the capabilities for (dynamic, asynchronous) automation
of these transactions. Overall, this restricts the application of these
specifications to classic scenarios involving a small number of
well-known authoritative documents over which a handful of questions is
repeatedly formulated. Moreover, the functional requirements for this
limited set of use cases are already fully supported by existing
solutions, such as the OpenID Connect ecosystem, which puts the need for
the new OpenID specifications into question. We therefore also look into
the non-functional advantages claimed by OpenID's new trust model: an
increase in control, privacy, and portability of personal information.
However, on none of these measures the new generation of specifications
clearly surpasses the existing one. Not only the technical choices limit
the capabilities of the EUDI framework; also the legislation itself
cannot accommodate the promise of self-sovereign identity. In
particular, we criticize the introduction of institutionalized trusted
lists, and discuss their economical and political risks. Their potential
to decline into an exclusory, recentralized ecosystem endangers the
vision of a user-oriented identity management in which individuals are
in charge. Instead, the consequences might severely restrict people in
what they can do with their personal information, and risk increased
linkability and monitoring. In anticipation of revisions to the EUDI
regulations, we suggest several technical alternatives that overcome
some of the issues with the architecture of OpenID. In particular,
OAuth's UMA extension and its A4DS profile, as well as their integration
in GNAP, are worth looking into. Future research into uniform query
(meta-)languages is needed to address the heterogeneity of attestations
and providers.
\end{abstract}
\vskip 3em

\renewcommand*\contentsname{Table of contents}
{
\setcounter{tocdepth}{2}
\tableofcontents
}

\section*{Executive summary}\label{sec-exec}

The EUDI regulation aims to create an EU-wide authentication framework
promoting self-sovereign identity, but are constructed around a limited
set of use cases. While the new generation of OpenID specifications
provides an architecture to implement these scenarios, existing
decentralized alternatives are equally capable of handling them. Lacking
a thorough conceptualization of (digital) identity, the EUDI regulation
and its OpenID architecture limit their possible applications, and
therefore fail to deliver the promised increase in control, privacy, and
portability of personal information.

Aggregating terminology from multiple international legal and technical
standards, we define (pseudonymous) \textbf{identity} as \emph{a subset
of measurable characteristics of an entity that, when taken together,
are sufficient to represent and distinguish that entity within a given
domain of application}. However, for practically all use cases, it
suffices to identify context-dependent \emph{roles} -- a form of
\textbf{partial identity}. Interpreting \emph{identification} as
\emph{the exchange of attributes as claims about a subject}, we
characterize the process of `\textbf{authentication}' as \emph{the
verification of authenticity of such claims} -- in particular their
\emph{origin} and \emph{integrity} -- based on \emph{trust} in some
\emph{authority} (i.e., the \textbf{issuer--holder--verifier} model). It
follows from these definitions that \emph{credentials are merely
certificates}: documents attesting to the truth of certain stated facts.

Based on these definitions, we find a mismatch between the
`paradigm-shifting' promises of EUDI and OpenID, and the actual
capabilities of their combined framework. The OpenID architecture, on
the one hand, is

\begin{itemize}
\tightlist
\item
  based on a history of software design that disregards best practices
  in internet security and interoperability;
\item
  limited to static credential types, and an inflexible,
  format-dependent query language; and
\item
  not applicable to dynamic or automated use cases.
\end{itemize}

Even within classic scenarios of credential exchange, the trust model
underlying OpenID4VCI and OpenID4VP offers no advantages over earlier
solutions, such as the OpenID Connect ecosystem:

\begin{itemize}
\tightlist
\item
  OpenID's wallet offers \emph{offline availability}, but does not
  increase \textbf{portability} of personal information in the usual
  \emph{Bring-Your-Own-Identity} sense, nor any extra control through
  \emph{informed consent}.
\item
  \emph{Selective disclosure} seems like a big step forward in user
  \textbf{control} over personal data; were it not that this form of
  \emph{data minimization} predominantly applies to wallet-based
  solutions, and is in fact not necessary in many other decentralized
  identity models.
\item
  Despite the change of interactions between issuer, holder, and
  verifier, \textbf{privacy} issues merely shift from the identity
  provider to the wallet provider -- or not at all, when following all
  the recommended security precautions.
\end{itemize}

The EUDI regulation, on the other hand, remains equally far from
achieving its promise of \emph{self-sovereign identity}. The use of
institutionalized \textbf{trusted lists} makes participation in the
identity market \emph{dependent on economical and political incentives}.
This risks a de facto \textbf{recentralization of digital identity} that
not only weakens the security and neutrality of the Web, but also makes
the framework vulnerable to \emph{vendor lock-ins}, \emph{data
correlation}, and \emph{government abuse}.

In anticipation of regulatory changes, we suggest alternatives that aim
at a broader interpretation of authentication and identity. The sound
foundation provided by W3C's decentralized identifiers (DID) and
verifiable credential model (VC) can be complemented by frameworks based
directly on OAuth. Ongoing research into the asynchronous and dynamic
capabilities of User-Managed Access (UMA), Authorization for Data Spaces
(A4DS), and the Grant Negotiation and Authorization Protocol (GNAP) are
prime candidates. Future research is still needed into uniform query
(meta-)languages, and support for more general attestation documents.

\begin{center}\rule{0.5\linewidth}{0.5pt}\end{center}

\newpage{}

\section*{Introduction}\label{introduction}
\addcontentsline{toc}{section}{Introduction}

The concept of \textbf{identity} plays a big role in the fields of
\emph{Data Spaces}\footnote{We use the term `\emph{data space}' here in
  its broadest sense, as any ecosystem in which participants exchange
  (provide and/or consume) data from heteregoneous sources through a
  unified interface (e.g., regardless of their format, physical
  location, or data model) {[}\citeproc{ref-curry:datapaces}{1}{]}. In
  particular, the term applies to abstractions within data management
  that shift the emphasis from up-front integration efforts with strong
  guarantees (e.g., consistency, durability), towards a paradigm based
  on co-existence and eventual semantic interoperability.} (DS) and
\emph{Identity and Access Management} (IAM): from the personal
information of an individual human being, over the information collected
about legal entities, to the processing of any of the former, either
through traditional methods or involving artificial
intelligence.\footnote{Many terms exist that define certain kinds of
  information about a person. Coming from different backgrounds, and
  using different terminology, it is often hard to determine how they
  relate. Most general is probably \textbf{personal information} (PI),
  which can be \emph{any} information related to an individual \emph{in
  any way} (e.g., one's favorite color, the number of steps it usually
  takes one to cross the market square). \textbf{Personal data} (PD),
  however, as defined in the GDPR {[}\citeproc{ref-eu:gdpr}{2}{]}, is
  personal information that ``is or might be directly or indirectly
  linked to {[}the subject{]} to whom such information relates''
  {[}\citeproc{ref-openid:connect}{3}{]}, ``in particular by reference
  to an identifier or to one or more factors specific to {[}their{]}
  identity'' {[}\citeproc{ref-eu:gdpr}{2}, art. 4{]}. It includes
  information that might not necessarily identify an individual
  \emph{directly} (i.e., on its own), but might nevertheless be
  \emph{indirectly} linked to them, when put in context, or when
  combined with other data. Examples include photographs, location data,
  and many technical identifiers (e.g., cookies, device identifiers,
  network addresses). \textbf{Personally Identifiable Information}
  (PII), a similar term from U.S. legislation, has a roughly equivalent
  definition, but is typically seen as somewhat narrower than personal
  data. It includes personal information that can be used to identify
  the individual to which that information relates
  {[}\citeproc{ref-openid:connect}{3}; \citeproc{ref-cnss:i4009}{4}{]},
  either alone (i.e., as a single data point) or ``when combined with
  other information that is linked or linkable to a specific
  individual'' {[}\citeproc{ref-nist:sp800-63}{5}{]}. Examples include
  contact details, account names, identification numbers -- but
  typically not the more technical identifiers, as present in personal
  data. A number of specific subsets of personal data -- and of PII --
  are defined in the GDPR, and together form \textbf{sensitive personal
  data}: information of a `private nature', requiring additional legal
  protection; e.g., ethnicity, sexual orientation, political opinions,
  religious beliefs, health-related data, and data that can be used for
  biometric identification. Lastly, the regulations in the scope of this
  paper introduce the term \textbf{Person Identification Data} (PID):
  the strictly smaller set of (sensitive) personal data ``enabling the
  identity of a natural or legal person \ldots{} to be established''
  {[}\citeproc{ref-eu:eidas}{6}{]}. It consists of the most fundamental
  personal details, in particular those issued and verified through
  official, authentic sources (e.g., one's name, nationality, date of
  birth).} For being so important, however, definitions of the term
`identity' are surprisingly rare in legal and technical documents
related to these fields.

In this paper, we look into the impact this has on the recent
developments around the European Digital Identity (EUDI) framework, and
its technical implementation through OpenID specifications. In
Section~\ref{sec-identity}, we present the state of the art on relevant
concepts, to establish a sound foundation to start from: we attempt to
aggregate a workable definition of the term `identity' and link it to
the concepts of anonymity and pseudonymity (\ref{sec-char}); we define
`authentication' and explain the trust model around certification
authorities (\ref{sec-trust}); and we look into the difference between
certificates and credentials (\ref{sec-attestations}).

Based on this understanding, we assess the technical capabilities of
OpenID's specifications in Section~\ref{sec-arch}: we discuss the design
choices inherited from OIDC (\ref{sec-oidc}), and point out the
limitations of OpenID4VCI (\ref{sec-vci}) and OpenID4VP (\ref{sec-vp}).
Given the lack of additional functionality, we try to find a rationale
for OpenID's new design by looking into its non-functional
characteristics Section~\ref{sec-paradigm}. We conclude, however, that
it does not offer any practical increase in portability
(\ref{sec-portability}), control (\ref{sec-control}) or privacy
(\ref{sec-privacy}).

Before offering a number of alternatives to OpenID's architecture in
Section~\ref{sec-alt}, we look into the relevant aspects of the EUDI
legislation itself in Section~\ref{sec-politics}: we explain the
workings of trusted lists (\ref{sec-lists}); discuss an example of the
effects of their institutionalization (\ref{sec-weakened}); and point
out the implications of the economical and political incentives they
give rise to (\ref{sec-recentral}).

\section{The identity of identity}\label{sec-identity}

Looking at the \emph{European Digital Identity} (EUDI) legislation
\emph{Regulation 2024/1183} {[}\citeproc{ref-eu:eudi}{7}{]}, as well as
its predecessor \emph{Regulation 910/2014} (eIDAS)
{[}\citeproc{ref-eu:eidas}{6}{]}, and its
\emph{Architecture and reference framework} (ARF)
{[}\citeproc{ref-eu:arf}{8}; \citeproc{ref-eu:toolbox}{9}{]}, it is
unclear precisely what is meant by the term `identity'; neither of them
provides an explicit definition. Likewise for many of the technical
specifications by Standards Development Organizations (SDOs),\footnote{SDOs
  mentioned in this section include: the World Wide Web Consortium
  (W3C); the OpenID Foundation (OpenID); the Organization for the
  Advancement of Structured Information Standards (OASIS); the
  International Organization for Standardization (ISO) and International
  Electrotechnical Commission (IEC), in particular their Joint Technical
  Committee 1 (JTC1); the National Institute of Standards and Technology
  (NIST); and the Committee on National Security Systems (CNSS).} which
play a major role in the framework's architecture: W3C's
\emph{Decentralized Identifiers} (DID) {[}\citeproc{ref-w3c:did}{10}{]},
\emph{Verifiable credentials data model} (VC)
{[}\citeproc{ref-w3c:vc}{11}{]}, {{[}\citeproc{ref-w3c:credapi}{12}{]}},
and \emph{Federated credential management} (FedCM)
{[}\citeproc{ref-w3c:fedcm}{13}{]}; OpenID's
\emph{OpenID for verifiable credential issuance} (OpenID4VCI)
{[}\citeproc{ref-openid:vci}{14}{]}, and
\emph{OpenID for verifiable presentations} (OpenID4VP)
{[}\citeproc{ref-openid:vp}{15}{]}. Older specifications, on which some
of these more recent standards are based, merely contain a brief
description of `identity': ``{[}a{]} set of attributes related to an
entity,'' in \emph{OpenID Connect} (OIDC)
{[}\citeproc{ref-openid:connect}{3}{]}; and ``{[}the{]} essence of an
entity \ldots{} described by one's characteristics,'' in OASIS's
\emph{Glossary for SAML} (SAML) {[}\citeproc{ref-oasis:glossary}{16}{]}
-- referencing the Merriam-Webster dictionary. These are far from
workable definitions for core specifications around this topic.

Note that the aim of this paper is not to find or attempt a universal,
one-size-fits-all definition of `identity'. Even more so than many other
concepts, its meaning heavily depends on the domain of governance
(legal, political, technical) and the sociocultural context. What we
point out is rather the \emph{lack of \textbf{any} definition} of the
term, neither given nor referenced, in the many texts pertaining to this
topic -- while still employing the term in other definitions. In
particular, when crossing from one domain to another -- e.g., from
legalislation to technical implementation -- lacking a (matching)
definition of this core concept poses the risk of operationalizing a
system that does not correspond to the original intentions.

We therefore turn to meta-standards and glossaries of those same SDOs.
Neither W3C's {{[}\citeproc{ref-w3c:glossary}{17}{]}}
{[}\citeproc{ref-w3c:glossary}{17}{]}, nor IETF's RFC 1983 (FYI 18)
({{[}\citeproc{ref-ietf:rfc1983}{18}{]}})
{[}\citeproc{ref-ietf:rfc1983}{18}{]}, or its RFC 7642
({{[}\citeproc{ref-ietf:rfc7642}{19}{]}})
{[}\citeproc{ref-ietf:rfc7642}{19}{]}, contain a definition of
`identity'. On the other hand, both IETF's RFC 4949 (FYI 36)
({{[}\citeproc{ref-ietf:rfc4949}{20}{]}})
{[}\citeproc{ref-ietf:rfc4949}{20}{]}, and RFC 6973
({{[}\citeproc{ref-ietf:rfc6973}{21}{]}})
{[}\citeproc{ref-ietf:rfc6973}{21}{]}, as well as OASIS's
\emph{Identity Metasystem Interoperability}
{[}\citeproc{ref-oasis:identity}{22}{]}, provide a somewhat detailed
definition of the term. Looking to expand our field of search, we find
several -- somewhat outdated -- documents giving an overview of the
state of the art (at the time), including the references mentioned
above: a `living paper' regularly updated at the Technical University of
Delft {[}\citeproc{ref-freehaven:terms}{23}{]}, the reports of the
\emph{Future of Identity in the Information Society} (FIDIS) project
{[}\citeproc{ref-fidis:journal}{24};
\citeproc{ref-fidis:book}{25}{]},\footnote{The FIDIS \emph{network of
  excellence} (NOE) was a \emph{European Research Area} (ERA) research
  consortium, funded by the European Commission's 6th Framework
  Programme for Research and Technological Development (FP6), with the
  aim to consolidate concepts and terminology related to identity, in
  order to foster interoperability of technologies and services towards
  a \emph{European Information Society} (EIS). Its conclusions were
  concretely taken up by, amongst others, its spiritual successor: the
  FP7 Specific Rargeted Research Project (STREP) \emph{Privacy and
  Identity Management for Community Services} (PICOS), focused on trust
  and security in information and communication technologies.} and the
extensive \emph{Security glossary}
{[}\citeproc{ref-wheeler:security}{26}{]} and \emph{Privacy glossary}
{[}\citeproc{ref-wheeler:privacy}{27}{]} of Wheeler \& Wheeler. In their
bibliographies, we discover a number of additional sources that provide
much more elaborate definitions of `identity' -- and numerous related
terms -- predominantly published by U.S. institutions: OECD/LEGAL/0491
{[}\citeproc{ref-oecd:govdi}{28}{]}, ISO/IEC 24760-1:2025
(\emph{A framework for identity management (part 1)})
{[}\citeproc{ref-iso:24760}{29}{]}, ISO/IEC 24765:2017
(\emph{Systems and software engineering vocabulary})
{[}\citeproc{ref-iso:24765}{30}{]}, NIST SP 800-63-4
({{[}\citeproc{ref-nist:sp800-63}{5}{]}})
{[}\citeproc{ref-nist:sp800-63}{5}{]}, NIST SP 800-103 (IPD)
(\emph{An ontology of identity credentials (part 1)})
{[}\citeproc{ref-nist:sp800-103}{31}{]}, NIST FIPS 200
(\emph{Minimum security requirements})
{[}\citeproc{ref-nist:fips200}{32}{]}, NIST FIPS 201-3
(\emph{Personal Identity Verification})
{[}\citeproc{ref-nist:fips201}{33}{]}, and CNSSI No. 4009
({{[}\citeproc{ref-cnss:i4009}{4}{]}})
{[}\citeproc{ref-cnss:i4009}{4}{]}.\footnote{Only while searching for
  these additional sources, we stumbled upon two white papers by OpenID,
  written to inform government officials, which contain a few of the
  same references {[}\citeproc{ref-openid:govcreds}{34};
  \citeproc{ref-openid:govdi}{35}{]}.} We discuss these in the following
section.

\subsection{Characteristics of (id)entities}\label{sec-char}

Of the handful of definitions we found, most are stated in terms of
\textbf{entities}. We paraphrase: an entity is anything that has
measureable \emph{attributes} (characteristics) by which it can be
\emph{represented} (named, described) and \emph{distinguished}
(recognized) in a relevant domain of applicability
{[}\citeproc{ref-ietf:rfc4949}{20}; \citeproc{ref-iso:24760}{29};
\citeproc{ref-iso:24765}{30}{]}. While most definitions of the term
\emph{attribute} turn out to be circular or self-referential,\footnote{Respectively,
  depending on \emph{identity}, or on -- not otherwise defined --
  variations of itself (e.g., feature, quality, characteristic)} we
manage to aggregate the following: an \textbf{attribute} is any unique
piece of information that \emph{associates} an entity with a measurable
\emph{value} of a particular \emph{type}
{[}\citeproc{ref-cnss:i4009}{4}; \citeproc{ref-eu:eidas}{6};
\citeproc{ref-ietf:rfc4949}{20}; \citeproc{ref-oecd:govdi}{28};
\citeproc{ref-iso:24765}{30}; \citeproc{ref-iso:27000}{36}{]}. An
\textbf{identity} is then \emph{any well-defined subset} of those
characteristics, that together suffice to \emph{actually} represent that
entity and uniquely distinguish it from any other entity within a
concrete context (domain) {[}\citeproc{ref-cnss:i4009}{4};
\citeproc{ref-nist:sp800-63}{5}; \citeproc{ref-ietf:rfc6973}{21};
\citeproc{ref-oecd:govdi}{28};
\citeproc{ref-nist:fips201}{33}{]}.\footnote{While this is a general
  definition, we use the term `identity' throughout this text to refer
  to `\emph{digital} identity' in particular: identity as captured,
  processed, and used within electronic system.}

Since any set of attributes, of which several entities share the same,
can be extended to a (minimal) distinguishing superset, we also call
sets of attributes in general \textbf{partial identities}
{[}\citeproc{ref-iso:24760}{29}{]}.\footnote{Pfitzmann express this in a
  witty way: ``An identity is any subset of attributes of an individual
  which identifies this individual within any set of individuals. So
  usually there is no such thing as `the identity', but several of
  them.'' {[}\citeproc{ref-freehaven:terms}{23}{]}} Interestingly,
partial identity immediately implies (partial\footnote{While identity
  can be qualified as partial, anonymity is typically not; though one
  could conceive of \emph{full} anonymity as the state in which the
  anonymity set contains all entities of the domain.})
\textbf{anonymity}: ``a state of an individual in which an observer
\ldots{} cannot identify {[}it{]} within a set of other individuals''
{[}\citeproc{ref-ietf:rfc6973}{21}{]}. Such a set of individuals, which
share ``the same attributes, making them indistinguishable from each
other {[}for a particular observer{]},'' is called an \emph{anonymity
set} {[}\citeproc{ref-ietf:rfc6973}{21}{]}. It is clear that any
nontrivial entity has many partial identities, and therefore many
different anonymity sets.\footnote{To count as anonymized data under EU
  legislation {[}\citeproc{ref-eu:gdpr}{2}{]}, the process obtaining the
  partial identity -- thereby removing actual identifiability -- from
  the proper one must be (reasonably) irreversible.}

A number of sources emphasize the distinction between identities and
\textbf{identifiers} {[}\citeproc{ref-oasis:glossary}{16};
\citeproc{ref-w3c:identity}{37}{]}, but they fail to point out the
precise difference: both are unique sets of characteristics. While
identifiers are often atomic strings (e.g., labels, serial numbers,
indexes), they can also be more complex or combined attribute structures
(e.g., names, addresses, bibliographic references)
{[}\citeproc{ref-cnss:i4009}{4}; \citeproc{ref-ietf:rfc4949}{20};
\citeproc{ref-iso:24765}{30}{]}. If anything, they are \emph{minimal}
(i.e., irreducible) identities, in which the removal of a single
attribute reduces it to a mere \emph{partial} identity. This aligns with
the recurring view that identifiers represent (other) \emph{identities}
-- larger (super)sets of attributes -- rather than \emph{entities}
{[}\citeproc{ref-openid:connect}{3}; \citeproc{ref-cnss:i4009}{4};
\citeproc{ref-iso:24760}{29}; \citeproc{ref-iso:24765}{30};
\citeproc{ref-nist:fips201}{33}; \citeproc{ref-w3c:identity}{37}{]}.

An entity potentially has multiple \emph{proper} (i.e., non-partial)
identities too, both in different contexts as within a single one. We
use the term \textbf{principal} (\emph{user}) to refer to the particular
identity, by which the active party\footnote{A party consists of ``one
  or more {[}entities{]} participating in some process or
  communication'' {[}\citeproc{ref-oasis:glossary}{16}{]}, which are
  ``considered to have {[}at least{]} some of the rights, powers, and
  duties of a natural person'' {[}\citeproc{ref-iso:24760}{29}{]}. They
  are natural or legal persons that have agency, and can therefore be
  held ``responsible for their actions and the actions of their agents''
  {[}\citeproc{ref-iso:24765}{30}{]}.} is known in the context of a
specific (user) session {[}\citeproc{ref-cnss:i4009}{4};
\citeproc{ref-ietf:rfc4949}{20}; \citeproc{ref-ietf:rfc6973}{21};
\citeproc{ref-iso:24760}{29}{]}.\footnote{A continuous period of
  transactions between an entity and an information system
  {[}\citeproc{ref-ietf:rfc4949}{20}{]}.} When a principal identity can
be ``associated with multiple interactions'' (with the same entity)
{[}\citeproc{ref-nist:sp800-63}{5}{]} -- because they contain at least
``the minimal identity information sufficient to \ldots{} establish it
as a link'' {[}\citeproc{ref-iso:24760}{29}{]} -- it is also called a
\textbf{persona} {[}\citeproc{ref-cnss:i4009}{4};
\citeproc{ref-w3c:vc}{11}{]}, or \textbf{pseudonym}.\footnote{Some
  sources hold pseudonyms and personas to conceal or protecting an
  entity's \emph{true identity} {[}\citeproc{ref-cnss:i4009}{4};
  \citeproc{ref-ietf:rfc4949}{20}{]} However, it is unclear what such a
  `\emph{true} identity' would be.} The latter term is typically used
when it concerns a \emph{minimal} persona: a privacy-preserving
\emph{identifier}, which does not let the verifier infer anything else
regarding the entity -- in particular not the identities by which the
entity is known to other parties {[}\citeproc{ref-openid:connect}{3};
\citeproc{ref-nist:sp800-63}{5};
\citeproc{ref-oasis:glossary}{16}{]}.\footnote{In EU law
  {[}\citeproc{ref-eu:gdpr}{2}{]}, data is called pseudonymized if on
  its own it can not be attributed to a specific subject entity, but --
  unlike anonymized data -- it can be (reasonably) reversed into a
  proper identity when in the possession of additional information.}

\subsection{On trust in authorities}\label{sec-trust}

The act of \emph{presenting} identity information to a system -- making
\textbf{claims} about the attributes of an entity (the
\textbf{subject})\footnote{In the context of identification, the subject
  is typically the principal itself, and the claims contained in the
  certificate are \emph{identity information}
  {[}\citeproc{ref-openid:vci}{14}; \citeproc{ref-ietf:rfc4949}{20};
  \citeproc{ref-iso:29100}{38}{]}. Because of this, the terms
  \emph{subject} and \emph{principal} are often used interchangeably
  {[}\citeproc{ref-cnss:i4009}{4}; \citeproc{ref-nist:sp800-63}{5};
  \citeproc{ref-iso:24760}{29}{]}.} -- and the system subsequently
\emph{validating}\footnote{To \textbf{approve} of something (e.g., data
  structures, relationships), typically implying that it satisfies
  certain \emph{requirements} (e.g., soundness, correctness,
  completeness, consistency), often in \emph{compliance} with some
  specification (e.g., standard, convention)
  {[}\citeproc{ref-ietf:rfc4949}{20}; \citeproc{ref-iso:24765}{30}{]}.
  In case of \emph{identification} (certificates), it involves checking
  whether all required attributes, needed to recognize an entity in a
  domain, are \emph{present}, have the \emph{correct syntax}, and are
  \emph{not expired} {[}\citeproc{ref-iso:24760}{29}{]}.} the collected
information in order to \emph{recognize} the represented entity -- as
(uniquely) distinct within a context -- is called
\textbf{identification} {[}\citeproc{ref-cnss:i4009}{4};
\citeproc{ref-nist:sp800-63}{5}; \citeproc{ref-ietf:rfc4949}{20};
\citeproc{ref-iso:24760}{29}{]}. \textbf{Authentication}, on the other
hand, is the more formal process of establishing a sufficient level of
\emph{assurance} in the \textbf{authenticity}\footnote{The property of
  ``being genuine and able to be verified and be trusted''
  {[}\citeproc{ref-cnss:i4009}{4}; \citeproc{ref-ietf:rfc4949}{20}{]},
  which -- in the case of data -- ensures confidence in its validity,
  its source (origin), and its transmission (integrity)
  {[}\citeproc{ref-nist:sp800-63}{5}{]}.}
{[}\citeproc{ref-openid:connect}{3}; \citeproc{ref-oasis:glossary}{16};
\citeproc{ref-oecd:govdi}{28}; \citeproc{ref-iso:24760}{29};
\citeproc{ref-nist:fips201}{33}{]} of an entity. It involves
\textbf{verifying}\footnote{Presenting \textbf{evidence} and checking it
  against previously corroborated information associated with it
  {[}\citeproc{ref-iso:24765}{30}; \citeproc{ref-nist:fips201}{33};
  \citeproc{ref-iso:29115}{39}{]}. It determines the \emph{truth} or
  \emph{accuracy} of an (atomic) statement (e.g., fact, claim, value,
  signature, integrity) {[}\citeproc{ref-openid:connect}{3};
  \citeproc{ref-ietf:rfc4949}{20}{]}, by \textbf{proving} the ``binding
  between the attribute and {[}the entity{]} for which it is claimed''
  {[}\citeproc{ref-ietf:rfc4949}{20}; \citeproc{ref-iso:29115}{39}{]}.}
the \emph{origin} (i.e., source) and \emph{integrity}\footnote{The
  ``constancy of and confidence in data values,'' i.e., the assurance
  that the data has not been altered (e.g., changed, destroyed, lost) in
  an unauthorized (or accidental) manner since it was
  created.{[}\citeproc{ref-ietf:rfc4949}{20};
  \citeproc{ref-nist:sp800-57}{40}; \citeproc{ref-nist:sp800-152}{41}{]}}
of information {[}\citeproc{ref-cnss:i4009}{4};
\citeproc{ref-eu:eudi}{7}; \citeproc{ref-ietf:rfc4949}{20};
\citeproc{ref-nist:sp800-57}{40}; \citeproc{ref-nist:sp800-175}{42}{]},
in order to determine the \emph{binding} between the presented claims
{[}\citeproc{ref-openid:connect}{3}; \citeproc{ref-nist:sp800-63}{5};
\citeproc{ref-oasis:glossary}{16}; \citeproc{ref-ietf:rfc1983}{18};
\citeproc{ref-ietf:rfc4949}{20}; \citeproc{ref-iso:24760}{29};
\citeproc{ref-nist:fips200}{32}; \citeproc{ref-iso:29115}{39};
\citeproc{ref-nist:sp800-57}{40}; \citeproc{ref-nist:sp800-175}{42};
\citeproc{ref-ietf:rfc2504}{43}{]}.\footnote{While authentication is
  therefore always a form of \emph{identity verfication}, it does not
  only apply to user identity: ``{[}it{]} may involve any type of
  attribute \ldots{} {[}claimed{]} on behalf of a subject or object''
  {[}\citeproc{ref-ietf:rfc4949}{20}{]}. A few sources claim that the
  term is therefore ambiguous, being either about \emph{identity}
  (`entity authentication'), about \emph{integrity} (`data
  authentication'), or about \emph{origin} (`source authentication')
  {[}\citeproc{ref-ietf:rfc4949}{20}; \citeproc{ref-nist:sp800-57}{40};
  \citeproc{ref-nist:sp800-175}{42}{]}. However, such a distinction
  misrepresents any of these `specific' processes as not including the
  other. To the contrary, it is impossible to check the origin of data
  without determining its integrity, and vice versa; and likewise, it is
  not possible to verify an identity without assuring its integrity and
  origin.} In the context of (user) identification, authentication
determines ``the degree to which the identity \ldots{} can be proved to
be the one claimed'' {[}\citeproc{ref-iso:24765}{30}{]}, typically by
ensuring that the entity controls (e.g., is, possesses, or knows) a
valid token (authenticator), bound to their account -- e.g., a private
key corresponding to a registered public key -- to demonstrate that they
are associated with that account {[}\citeproc{ref-nist:sp800-63}{5};
\citeproc{ref-w3c:webauthn}{44}{]}.\footnote{Note that successful
  authentication does not imply that the verifier knows the
  \emph{natural} identity of a user. It only guarantees the relying
  party that, when multiple sessions involve the same authenticated
  attributes, these sessions pertain to the same principal. The natural
  person behind that principal might, however, not always be the same
  {[}\citeproc{ref-w3c:webauthn}{44}{]}.}

Assurance about the authenticity of information can be based on trust in
third-party \textbf{(certification) authorities} (CAs) -- the archetypal
\emph{trust service provider} {[}\citeproc{ref-eu:eidas}{6}{]} -- that
vouch for its integrity and accuracy. These trusted authorities are
often the primary or \emph{authoritative sources}, who manage the life
cycle of the information, keeping it accurate and up-to-date
{[}\citeproc{ref-iso:29115}{39}{]};\footnote{A life cycle consists of
  the different stages involved in the management of information: its
  collection, initial authentication (\emph{proofing}), and registration
  (storage) -- in case of a \emph{digital identity lifecycle} also
  together called \emph{enrollment} {[}\citeproc{ref-oecd:govdi}{28}{]}
  -- as well as its issuance, revocation, expiration, preservation, and
  termination {[}\citeproc{ref-cnss:i4009}{4};
  \citeproc{ref-nist:sp800-63}{5}; \citeproc{ref-ietf:rfc4949}{20};
  \citeproc{ref-iso:24760}{29}{]}.} as well as the \emph{issuers}, who
generate (signed) documents (\textbf{certificates}) asserting the
information, and provide them to principals
{[}\citeproc{ref-openid:connect}{3}; \citeproc{ref-nist:sp800-63}{5};
\citeproc{ref-eu:eidas}{6}; \citeproc{ref-w3c:vc}{11};
\citeproc{ref-openid:vci}{14}; \citeproc{ref-openid:vp}{15};
\citeproc{ref-ietf:rfc4949}{20}; \citeproc{ref-iso:24760}{29};
\citeproc{ref-iso:24765}{30}; \citeproc{ref-nist:fips201}{33}{]}. When
the asserted information is about the principal themself, we also call
such authorities \textbf{credential service providers} (CSPs) or
\emph{identity providers} (IDPs), and the certificates they issue
\textbf{(verifiable) credentials} -- or, to use the EUDI term,
\emph{(electronic) attestations of attributes} (EAAs)
{[}\citeproc{ref-eu:eudi}{7}; \citeproc{ref-eu:arf}{8};
\citeproc{ref-oasis:glossary}{16}; \citeproc{ref-ietf:rfc6973}{21};
\citeproc{ref-oecd:govdi}{28};
\citeproc{ref-iso:24760}{29}{]}.\footnote{In Section~\ref{sec-politics},
  we discuss different kinds of EAAs, and go deeper into the nature of
  trust services and their providers.}

Having received a certificate, its \textbf{holder} -- the principal to
whom it was issued and who controls it
{[}\citeproc{ref-nist:sp800-103}{31}{]} -- can transmit it to third
parties as \textbf{(verifiable) presentations}
{[}\citeproc{ref-nist:sp800-63}{5}; \citeproc{ref-w3c:vc}{11};
\citeproc{ref-openid:vci}{14}; \citeproc{ref-openid:vp}{15}{]}: a
selection of assertions (derived) from one or more certificates
{[}\citeproc{ref-w3c:vc}{11}; \citeproc{ref-openid:vp}{15}{]}, typically
bound to (a key of) the holder -- to prove their legitimate possession
of the certificate in question {[}\citeproc{ref-openid:vp}{15}{]}. The
third party service provider, who receives a (verifiable) presentation
from a principal, and performs verification to assess its authenticity,
is called the \textbf{verifier} {[}\citeproc{ref-w3c:vc}{11};
\citeproc{ref-openid:vci}{14}; \citeproc{ref-openid:vp}{15};
\citeproc{ref-iso:24760}{29}{]}, or \textbf{relying party} -- because it
trusts and \emph{relies} on a system of digital identity solutions to
confirm the validity of the assertions {[}\citeproc{ref-w3c:vc}{11};
\citeproc{ref-oasis:glossary}{16}; \citeproc{ref-ietf:rfc4949}{20};
\citeproc{ref-oecd:govdi}{28}; \citeproc{ref-iso:24760}{29}{]}.

\subsection{An anatomy of attestations}\label{sec-attestations}

Certificates typically consist of two distinct types of information.
First, they include the actual content, as one or more
\textbf{assertions}: claims made by the issuer about the \emph{subject},
associating it with certain attribute values
{[}\citeproc{ref-openid:connect}{3}; \citeproc{ref-w3c:vc}{11};
\citeproc{ref-openid:vci}{14}; \citeproc{ref-oasis:glossary}{16};
\citeproc{ref-oasis:saml}{45}{]}. Second, it contains
\textbf{attestation} information: evidential (meta-)statements that
describe the \textbf{authentication context}, consisting of all the
``information that the relying party can require before it makes {[}a{]}
decision with respect to an authentication''
{[}\citeproc{ref-openid:connect}{3}{]}. This includes the verification
and revocation methods, level of assurance, security characteristics,
and all the (cryptographic) \textbf{verification factors} needed to
``verify the security of cryptographically protected information''
{[}\citeproc{ref-w3c:vc}{11}{]}. These factors bind the asserted claims
together, to the issuer and -- in case of credentials -- to the
principal {[}\citeproc{ref-cnss:i4009}{4}{]}.

It is often assumed that ``the holder of a credential \ldots{}
presenting the claims {[}to{]} the verifier is (controlled by) the
subject of the claims'' {[}\citeproc{ref-openid:security}{46}{]}. Some
sources note certain exceptions, in which ``claims in a credential can
be about different subjects'' {[}\citeproc{ref-w3c:vc}{11}{]} -- not
limited to assertions about the principal presenting them -- or in which
the holder is not a subject of the claims at all (e.g., a parent
presenting their child's birth certificate)
{[}\citeproc{ref-openid:security}{46}{]}. These cases are in essence an
abstraction from credentials towards certificates in general. They often
lack technical support, though, because relying parties typically
require information about the user they actually interact with during a
session. Without this guarantee, malicious parties can abuse credentials
(e.g., obtained through a data leak or other security breach) to
\emph{impersonate} someone else
{[}\citeproc{ref-openid:security}{46}{]}. These scenarios underlie the
idea of \textbf{\emph{`legitimate'} holders} of a credential, often --
but not always -- the subject.

Credentials are therefore often tied to their (legitimate) holder,
through a mechanism called \textbf{holder binding}: the addition of
holder-related verification factors. This additional attestation
information serves as evidence, to be corroborated with information
known to be in control of the legitimate holder. It can be something the
holder \emph{has} or \emph{knows}, or something they \emph{are} or
(typically) \emph{do} {[}\citeproc{ref-iso:29115}{39}{]}, including
traditional accounts (e.g., password-based), cryptographic material
(e.g., keys), biometric information (e.g., fingerprints), or any other
claim linking the credential in question to information (i.e., another
credential) of which the holder's identity has already been established
{[}\citeproc{ref-openid:vci}{14}; \citeproc{ref-openid:security}{46}{]}.
Nevertheless, even OpenID -- whose specifications indeed preclude
scenarios in which holder and subject differ (cf. Section~\ref{sec-vci})
-- admits that it is ``important to distinguish between the information
that the credential holds (about the subject) and the information that
the credential is bound to (about the holder)''
{[}\citeproc{ref-openid:security}{46}{]}. This makes it hard to maintain
a useful distinction between credentials and certificates -- not unlike
between identities and identifiers.

The distinction becomes even more blurry when considering that
identification does not necessarily need to be unique. In a large amount
of use cases, it suffices that the relying party knows the principal's
\emph{partial} identity.\footnote{One classical example is the age
  requirement for buying alcohol. A minimal credential, that still
  provides sufficent assurance to an online shop, could consist of an
  age -- or merely a `yes' or `no' -- combined with a verification
  factor, and signed by a government institution. The only necessary
  information is the binding of the holder (i.e., the buyer) with the
  partial identity ``someone of legal age'' -- an identity that could be
  taken up by the majority of people. Solely based on these credentials,
  it is impossible for the shop to know whether multiple successful
  checkouts were made by the same buyer or not. Without extra
  information (e.g., customer accounts), from the perspective of the
  shop, the world consists of only two identities: one over the legal
  age, and one not.} This is also apparent in the strong increase of
\textbf{role-based access control} mechanisms over (purely)
\emph{identity-based} ones, amongst others in many cloud services (e.g.,
Google Cloud, Amazon Web Services).\footnote{In the role-based paradigm,
  entities are identified by their \textbf{roles}: formal placeholders
  for the \emph{functional positions} in which they can participate in
  interactions {[}\citeproc{ref-ietf:rfc4949}{20};
  \citeproc{ref-iso:24765}{30}{]}. Roles are characterized by
  constraints on their behavior expected in a specific situation, i.e.,
  a set of \emph{connected actions} expressing
  \emph{situation-dependent} attributes
  {[}\citeproc{ref-freehaven:terms}{23}{]}. They are typically expressed
  as a pre-established collection of applicable policy rules
  {[}\citeproc{ref-ietf:rfc4949}{20}; \citeproc{ref-iso:24765}{30}{]}:
  ``when you grant a role to a principal, you give that principal all of
  the permissions in that role'' {[}\citeproc{ref-gc:iam}{47}{]}.}
Entities are then issued a \emph{role certificate} -- in place of
individual identifiers {[}\citeproc{ref-ietf:rfc7642}{19}{]} -- which
only asserts that the entity is ``{[}a{]} member of the set of
{[}entities{]} that have identities that are assigned to the same role''
{[}\citeproc{ref-ietf:rfc4949}{20}{]}. One could even go so far as to
say that \emph{roles \textbf{are} (partial) identities}, i.e., sets of
attributes shared by different entities
{[}\citeproc{ref-aws:iam}{48}{]}. Which roles or (partial) identities
are distinguished depends entirely on ``the context of a function
delivered by a particular application''
{[}\citeproc{ref-auth0:glossary}{49}{]}.

When we take all these nuances into account, \emph{the subject becomes
of less and less importance}. As a concrete example, take the sale of a
property. To be able to sell it, the owners should at least provide a
credential (i.e., the deed) attesting that they -- as a subject -- are
indeed it's legitimate proprietors. A number of other relevant
documents, however, should -- while definitely bound to the property --
not necessarily mention them (e.g., energy ratings, attestations of soil
composition). Both the deed and the other documents will have to be
transferred to the notary, their information verified at the relevant
institutions, and subsequently processed into a bill of sale, signed by
all parties involved, and passed onto the buyer. While the presence or
absence of the holder--subject thus forms a single (theoretical)
difference between credentials and certificates, the (practical)
similarities are much more numerous. We therefore conclude that -- at
least from a practical perspective -- there is \emph{no useful
distinction between credentials and (other) certificates}. Both are a
``document attesting to the truth of certain stated facts''
{[}\citeproc{ref-nist:fips201}{33}{]}.

\section{OpenID's architecture for EUDI}\label{sec-arch}

We now consider how this more elaborate understanding of the involved
concepts -- in particular the realization that identity and credentials
can be \emph{any (certified) data} -- impacts the EUDI framework and its
OpenID architecture. We start with a brief technical refresher.

The basis of the entire OpenID ecosystem is the OAuth 2.0 Authorization
Framework {[}\citeproc{ref-ietf:rfc6749}{50}{]}, which moves the
responsibility for access control from the \emph{resource server} (RS)
to a separate \emph{authorization server} (AS). The latter provides
\emph{clients} (applications) of the former with \textbf{access tokens},
which enable access to protected resources, in exchange for a variety of
\textbf{grants}: credentials that represent the resource owner's
approval of the requested access (e.g., client-specific credentials,
interactively obtained codes). This exchange happens at the (singular)
\textbf{token endpoint}, where the client \emph{authenticates} itself,
presents its grant, and requests a certain resource scope
{[}\citeproc{ref-ietf:rfc6749}{50}{]}. The OAuth 2.0 authorization
server is therefore both an \emph{authority} and a \emph{relying party}:
it issues tokens, but verifies grants. In terms of identity, it
exchanges a client's \emph{proper} identity -- linked to the grant --
for a \emph{partial} one: the token, which merely asserts that the
client has a certain authorization. Note that the input (grants) can
thus be a lot more complex than the output (tokens). This is also
apparant in the variety of grant-related extensions to the
protocol.\footnote{While the extensions related to tokens are at least
  as numerous, they mostly concern security enhancements, or additional
  JSON Web Token claims.}

\subsection{OpenID Connect}\label{sec-oidc}

All OpenID's architectures (OIDC, OpenID4VCI, OpenID4VP) are layered on
top of this OAuth 2.0 design, reusing its \emph{authorization}
primitives (flows, endpoints, tokens etc.) for \emph{authentication}
purposes.\footnote{In doing so, OIDC -- and every other OpenID
  specification -- breaks one of the central characteristics of OAuth
  2.0: the \emph{separation of orthogonal concerns}. An OIDC server
  manages \emph{both} the identity data itself \emph{and} the
  permissions to access that data. This results in a less modular, less
  extensible architecture, and creates a lot of confusion around the
  semantics of core OAuth 2.0 concepts (e.g., tokens, audience,
  revocation). Two pertinent implications of this are implementations
  that (1) employ the OIDC access token to access resources other than
  the userinfo endpoint; or (2) mint identity tokens for other audiences
  than the client. Moreover, as a consequence, such implementations do
  not always comply with the necessary security measures for OIDC's
  threat model.} OIDC repurposes OAuth 2.0's interactive authorization
code flow,\footnote{\label{fn-oidc-flow}We do not go into the other
  flows specified in OIDC, since they are based on the insecure implicit
  flow from OAuth 2.0. This flow has since a long time been found
  vulnerable to several attack vectors, and was therefore deprecated by
  RFC 9700 (BCP 240) (\emph{OAuth 2.0 security BCP})
  {[}\citeproc{ref-ietf:rfc9700}{51}{]} and removed in the draft of
  \emph{OAuth 2.1} {[}\citeproc{ref-ietf:oauth21}{52}{]}.} in which the
principal authenticates themselves, and returns an \textbf{identity
token} (ID token) -- containing a \emph{subject identifier} -- and
optionally an access token, which enables the client to retrieve
(additional) identity information from the identity provider's
\textbf{userinfo endpoint} {[}\citeproc{ref-openid:connect}{3}{]}.

By issuing ID tokens from the same token endpoint as access tokens,
however, OIDC had to shoehorn much more complex information -- \emph{any
form of identity data} -- into an API that was designed for a much
simpler (yes-or-no) output. While the extra userinfo endpoint seems to
accommodate for this added complexity, its interface only allows for
\emph{a flat key-value mapping} (from attribute names to JSON values).
Moreover, specific claims from this mapping are requested using a
non-standardized query language -- best described as a custom version of
JSONPath {[}\citeproc{ref-ietf:rfc9535}{53}{]}, or a primitive flavor of
JSON Schema {[}\citeproc{ref-openid:connect}{3}{]}. Even in their
standardized form, these JSON-based languages are structurally coupled
to the JSON value tree, and thus to a particular credential format,
rather than to the semantic content of the attestation. This limits the
(re)usability of a query, since it ``imposes an implicit reliance on
\ldots{} the issuer's local context, such as language and culture''
{[}\citeproc{ref-idlab:queryvc}{54}{]}. It is no surprise then, that --
despite the extensibility of OIDC's set of claims -- implementations
typically only rely on the subject identifier, or at most a limited
selection of standard claims (name, email, address, phone number, birth
date, gender).

\subsection{OpenID for Verifiable Credential Issuance}\label{sec-vci}

In OpenID4VCI, identity information is no longer provided as an ID
token, but the essence remains the same: clients can request information
as a credential using an OAuth 2.0 flow,\footnote{\label{fn-vci-flow}Note
  that even in this recent specification, OpenID does not completely bar
  the insecure flows from OAuth 2.0 (cf.~Footnote \ref{fn-oidc-flow}).
  This is especially surprising since they refer to the deprecation of
  these flows by RFC 9700 (BCP 240)
  {[}\citeproc{ref-ietf:rfc9700}{51}{]} in their own OpenID4VCI
  \emph{Security Considerations} {[}\citeproc{ref-openid:vci}{14, Sec.
  13.2}{]}, OpenID
  \emph{OpenID4VC high assurance interoperability profile}
  {[}\citeproc{ref-openid:haip}{55}{]}, and OpenID
  {{[}\citeproc{ref-openid:fapi}{56}{]}}
  {[}\citeproc{ref-openid:fapi}{56}{]}.} and access it at the credential
endpoint -- replacing the userinfo one
{[}\citeproc{ref-openid:vci}{14}{]}. Nevertheless, EUDI implementers
claim that its design has ``several essential benefits'' for the
ecosystem: it enhances trust, security, privacy, control, compliance,
and interoperability. As underlying reason for these benefits, they
state that OpenID4VCI combines the well-known flows and straightforward
user experience of OIDC with the `new' digital proofs technology of
verifiable credentials -- which they claim to be harder to fake or alter
{[}\citeproc{ref-igrant:eudi}{57}{]}.

Practical attempts tell a different story, though. VCs indeed provide a
semantically richer alignment of credentials, but they are not
inherently more expressive, nor more secure, than classic OIDC tokens.
To the contrary, the OpenID4VCI specification itself lists JSON Web
Tokens (JWTs) with JSON Web Signatures (JWS)
{[}\citeproc{ref-ietf:rfc7519}{58}; \citeproc{ref-ietf:rfc7515}{59}{]}
-- similar to an OIDC token -- as a possible serialization of VCs
{[}\citeproc{ref-openid:vci}{14}{]}.

Moreover, while the query mechanics got a slight upgrade, it remains
focused on a narrow concept of credentials. Issuers specify a number of
preset \textbf{credential configurations} in their metadata: particular
combinations of a \textbf{credential type} with a \textbf{credential
format} (e.g., ``a driver's license in ISO's mdoc format'', ``a
university degree in SD-JWT format''). Concrete credentials are then the
result of applying a credential configuration to a dataset of identity
information {[}\citeproc{ref-openid:vci}{14}{]}. Within these
configurations, issuers can provide clients a decent amount of
flexibility regarding the \emph{packaging} (e.g., format parameters,
cryptographic algorithms); yet only a single parameter addresses the
actual \emph{content} of the credential. The entire semantics,
differentiating one credential (configuration) from another, must be
expressed in a single string -- the \emph{credential type}. This is only
practically feasible when the offered configurations are static, and
limited in number.

Using \textbf{claim descriptions} (sets of \emph{claims path pointers}),
clients can select a limited number of claims out of a larger credential
(configuration) {[}\citeproc{ref-openid:vci}{14}{]}, but this only adds
the ability to request \emph{subsets} of the predefined credential types
offered by the issuer. As such, it is no improvement over OIDC's custom
flavor of JSONPath or JSON Schema (cf.~supra) -- yet is not
interoperable with any of them, nor compatible with any existing OpenID
client. Moreover, this approach makes the semantics of a credential
dependent on the structure of the format -- and assumes that the client
already knows this structure. OpenID4VCI is thus only suited for issuers
offering a select assortment of distinct bundles of information, with a
\emph{fixed semantics} agreed upon out-of-band.

The design of OpenID4VCI also precludes the issuance of credentials with
subjects other than the principal. In scenarios where the client is not
yet known -- i.e., all cases except active offers by the issuer or
updates with refresh tokens -- \emph{the principal must (interactively)
identify themself}. Since every other step of the protocol is based on
\emph{subject-agnostic} credential (configuration) identifiers, there is
no other way for the issuer to know about which subject a credential is
requested, i.e., to which dataset to apply the credential configuration.
Not only does this severely \emph{limit the kind of credentials} that
can be issued through OpenID4VCI; the need for interaction also
\emph{complicates the automation} of credential issuance. Taken
together, the limitations described in this section lead us to conclude
that OpenID4VCI is mainly targeted at the issuance of a handful of
specific credentials, actively offered to (the client of) a known
principal.\footnote{This conclusion is reinforced by the protocol's lack
  of separation between a resource server and an authorization server.
  While the specification explicitly states that ``the Credential Issuer
  acts as {[}a{]} Resource Server {[}which{]} \emph{might} also act as
  an Authorization Server.''{[}\citeproc{ref-openid:vci}{14}{]} -- thus
  allowing for (multiple) authorization servers separate from the
  credential issuer -- the practicalities of the protocol prevent it
  from being actually feasible. The authorization servers supported by
  the issuer are merely declared as a flat array in its metadata.
  Without an offer, there is no way for clients to know which one to
  request a token from; nor is there a mechanism specified for dynamic
  discovery at the credential endpoint (e.g., as part of the error
  response when unauthenticated). When not (only) relying on credential
  offers, the issuer itself has therefore to be the (only) authorization
  server -- thereby severely limiting the flexibility of the framework.}

\subsection{OpenID for Verifiable Credential
Presentations}\label{sec-vp}

Credentials issued via OpenID4VCI are not meant to be requested by -- or
transmitted to -- verifiers themselves. Instead, the EUDI framework
defines an extra intermediary trust service: a digital \textbf{wallet},
which requests credentials from issuers, and presents them to verifiers
{[}\citeproc{ref-eu:eudi}{7}{]}. OpenID specifies the design of these
wallets in OpenID4VP {[}\citeproc{ref-openid:vp}{15}{]}. Unsurprisingly,
it suffers from many of the same issues as OIDC and OpenID4VCI --
without offering any practical innovations that would warrant a new
specification.

Even more than in those other specifications, the architecture of
OpenID4VP forces the wallet to be both authorization server and resource
server. This precludes more flexible scenarios, in which these roles are
federated or otherwise decentralized. Moreover, while it claims to
provide enhanced security to the verification mechanism -- and thereby
to amplify the credibility of digital authentication procedures
{[}\citeproc{ref-igrant:eudi}{57}{]} -- the specification seems to
double down on a variant of the insecure OAuth 2.0 implicit
authorization flow, which has since a long time been deprecated
{[}\citeproc{ref-ietf:rfc9700}{51}{]}, and is no longer an option in
OAuth 2.1 {[}\citeproc{ref-ietf:oauth21}{52}{]}
(cf.~Footnotes \ref{fn-oidc-flow} and \ref{fn-vci-flow}).

Similar to the OIDC flow, a successful OpenID4VP request results in a
response containing one or more credentials -- now called VP tokens
instead of ID tokens. To request and construct these (verifiable)
presentations from the credentials available to the wallet, OpenID4VP
includes a custom Digital Credentials Query Language (DCQL). There is
not much more to this language than some metadata around the JSONPath
style \emph{claims descriptions} of OpenID4VCI: sets of \emph{claims
path pointers} indicating the desired claims by their structural
location in a known credential type (predefined out-of-band). Rather
than achieving the claimed `comprehensive interoperability'
{[}\citeproc{ref-igrant:eudi}{57}{]}, the custom interfaces of the
specification thus preclude a straightforward integration between
conventional and contemporary OpenID technologies -- let alone other
identity management frameworks.

\section{A paradigm shift that never was}\label{sec-paradigm}

Given the strong parallel between OpenID4VP and OpenID4VCI, highlighted
in the previous section, one must wonder about the reasons for having
two separate specifications that regulate almost identical flows of
information. After all, (verifiable) credentials \emph{are} (verifiable)
presentations themselves. In OIDC, for example, relying parties get the
identity token (i.e., presentation) of a principal directly from the
latter's issuer.

The similarity becomes even more apparent when taking into account
\emph{self-issued credentials}. Extending OIDC, the
\emph{Self-issued OpenID provider} (SIOPv2)
{[}\citeproc{ref-openid:siop}{60}{]} specification defines an OpenID
provider (i.e., issuer) controlled by the principal -- either in the
cloud or on their device, similar to a wallet -- such that ``the
{[}principal{]} becomes the issuer of identity information {[}i.e.,
tokens{]}, signed with keys under {[}their{]} control {[}in order to{]}
present self-attested claims directly to the {[}relying party{]}''
{[}\citeproc{ref-openid:siop}{60}{]}.\footnote{Note that the SIOPv2
  specification explicitly aligns itself with self-signed certificates
  such as W3C presentations with self-asserted claims
  {[}\citeproc{ref-w3c:vc}{11}; \citeproc{ref-openid:siop}{60}{]}.
  However, SIOPv2 tokens -- like any OpenID tokens -- are limited to
  assertions about the principal themself, while the W3C model
  explicitly states that ``self-asserted claims \ldots{} are
  \textbf{\emph{not}} limited {[}to{]} statements about the holder''
  {[}\citeproc{ref-w3c:vc}{11}{]}.} Though the relying party's trust
relationship in SIOP is directly with the principal, rather than with a
third-party issuer, from a technical perspective OIDC and SIOPv2 are
almost identical: only the equality of the \texttt{sub} and \texttt{iss}
claims indicates the difference.\footnote{Interestingly, the SIOPv2
  specification states that, unlike OIDC assertions, the information
  included in a SIOPv2 token is -- due to its self-asserted,
  self-attested nature -- \textbf{non-verifiable}: allegedly, signatures
  made by the principal themself can not be used to validate the origin
  of the signed information, because the principal ``does not have the
  legal, reputational trust of a traditional {[}issuer{]}''
  {[}\citeproc{ref-openid:siop}{60}{]}. We beg to differ: in case of
  self-issued identity presentations, the origin \emph{is} the
  principal; and by checking their signature this origin is verified.
  Moreover, nothing precludes a principal and a relying party to have a
  \emph{legal} or \emph{reputational} trust relationship. Most
  importantly, however, trust has no fundamental need for such
  `rational' qualifiers: as soon as a relying party trusts the principal
  -- for whatever reason -- the latter's signature will per definition
  be sufficient for the former.} However, while OIDC and SIOPv2 employ a
single, uniform protocol (OIDC), on top of which self-issuance is made
possible, OpenID4VCI and OpenID4VP are -- while functionally the same --
specified separately. The question therefore remains: what are the
advantages of such a double tiered framework?

In a white paper, titled \emph{}
{[}\citeproc{ref-openid:trustmodel}{61}{]}, OpenID calls its own
architecture constitutive of a \emph{paradigm shift}, driven by an
evolution in \emph{user-centricity}: the principal is put ``in the
center of the exchange between the verifier and the credential issuer''
{[}\citeproc{ref-openid:trustmodel}{61}{]},\footnote{Note that -- while
  other readings are possible -- the lack of a comma in the phrase ``the
  center of the exchange between the verifier and the credential
  issuer'' prompts a traditionally schooled reader to interpret this
  sentence as pertaining to an exchange between \textbf{two} parties
  (i.e., verifier and issuer). The principle may be at the center of it,
  but is -- under this interpretation -- still not a main `partner'.}
granting them more \textbf{control}, \textbf{privacy}, and
\textbf{portability} of their identity information. This `big shift'
must be understood against the background of `traditional'
\textbf{federated models}.\footnote{A systems model in which parties
  collectively agree upon a division of the entire system into a set of
  independent `local' domains, \emph{within} each of which decisions are
  made autonomously according to their own interests, while interactions
  \emph{between} them are governed by a `global', mutually agreed
  regulation {[}\citeproc{ref-serrano:federated}{62};
  \citeproc{ref-narayanan:decentralized}{63}{]}.} In such systems, an
issuer -- in a trust relationship with the verifier -- provides
credentials \emph{just-in-time}, i.e., each time a principal requires
one for interacting with a relying party. OpenID's new architecture
figures in the trend of \textbf{decentralization},\footnote{A systems
  model in which (some) activities or processes are not controled by a
  single, central authority
  {[}\citeproc{ref-johnson:decentralized}{64}{]}. Rather, independent
  decisions are made locally and autonomously by multiple parties
  (called \emph{peers}), directed towards their own individual -- and
  possibly conflicting -- goals
  {[}\citeproc{ref-khare:decentralized}{65}{]}.} which they
\emph{contrast} with federation. This is surprising, since federation is
actually \emph{one of two forms of decentralization} -- the other one
being distributed (peer-to-peer) models
{[}\citeproc{ref-narayanan:decentralized}{63}{]}.\footnote{Taking the
  distributed models as the opposite of fully centralized models, one
  could think of federation as covering the entire spectrum between --
  but excluding -- those extremes. As such, a federated system is one
  that has \emph{multiple centers} instead of one: a ``distributed
  network with each node \ldots{} being a centralized network,'' as it
  were {[}\citeproc{ref-peeters:decentralized}{66}{]}.} Since the OpenID
ecosystem is not a fully distributed one, we look into the benefits they
attribute to their `beyond federated' decentralized approach, and how
these benefits contribute to the claimed increase in \emph{control},
\emph{privacy}, and \emph{portability}.

\subsection{Portability}\label{sec-portability}

As OpenID correctly states in their white paper, \emph{portability} of
identity information increases in a decentralized system, because
principals can use \emph{their \textbf{own} identifier(s)} at issuers
and verifiers, instead of one namespaced to a specific third-party
issuer and assigned to them {[}\citeproc{ref-openid:trustmodel}{61}{]}.
However, this \textbf{Bring Your Own Identity} (BYOI) concept is already
a cornerstone of `traditional' \emph{federated} models -- largely
popularized by OIDC itself. The question therefore remains: how does
OpenID move beyond this in their new architecture, and what do users
gain by it?

According to OpenID's white paper, the increase in portability consists
of the principal's ability to ``control their relationship with the
verifiers \emph{independent} from third party {[}issuers'{]} decisions
or lifespan,'' and therefore to ``present credentials to the relying
parties who do \emph{not have a federated relationship} with the
credential issuer'' {[}\citeproc{ref-openid:trustmodel}{61}{]}. This is
a strong claim, for which the white paper provides no support, nor any
use case in which it would be required. In fact, in the same text,
OpenID writes that ``verifiers need to trust the respective credential
issuer,'' and that the establishment of such cross-domain,
inter-organizational trust will require ``regulatory or contractual
relationships on top of technical interoperability''
{[}\citeproc{ref-openid:trustmodel}{61}{]}. It is precisely this legally
and technically supported trust relationship which constitutes a
federation; and OpenID's new architecture relies on it just as much as
any other federated model.

If anything, the OpenID4VP specification increases portability in a more
literal sense: storing credentials in a wallet may ensure their
\textbf{offline availability} in case of technical problems at the
issuer's side, or -- with an on-device wallet -- in case of general
internet failure. Then again, such functionality could also be
implemented through existing specifications for asynchronous federation,
e.g., OIDC's Claims Aggregation extension
{[}\citeproc{ref-openid:aggregation}{67}{]}. None of the aspects of
increased portability discussed in this section can therefore truly be
attributed to OpenID's new design.

\subsection{Control}\label{sec-control}

According to OpenID's white paper, from a \emph{control} perspective
`decentralization' means ``not depending on one single body controlling
\ldots{} the ecosystem,'' and thus enabling principals and other parties
to make critical decisions, e.g., ``from \emph{which {[}issuer{]}} to
obtain what credential,'' and ``\emph{when} to disclose \emph{which
credential} to \emph{which verifier}''
{[}\citeproc{ref-openid:trustmodel}{61}{]}. Apart from the promised
increase in portability (cf.~supra), the specifications seem to rely
predominantly on \textbf{informed consent} and \textbf{selective
disclosure} to check this box.

\emph{Informed consent} is indeed a cornerstone of privacy-aware
technology. Verifiers and wallets should make sure that the context of a
request -- including a sufficiently specific purpose -- is clear to the
principal, and should obtain the latter's consent -- through explicit
interaction -- before disclosing information
{[}\citeproc{ref-openid:vp}{15}{]}. Again, however, this is nothing new
-- OIDC already stresses its importance. Neither does it answer the
question concerning the split architecture, since both OpenID4VP
\emph{and} OpenID4VCI require consent {[}\citeproc{ref-openid:vci}{14};
\citeproc{ref-openid:vp}{15}{]} -- thereby redundantly overloading the
user experience worse, contrary to their own claims (cf.
Section~\ref{sec-vci}).

\emph{Selective disclosure} is an interesting feature. This data
minimization technique -- supported by multiple credential formats --
enables principals to select specific claims from a credential in their
wallet, creating a presentation that only discloses the selected
information, without revealing the rest of the credential to the
verifier {[}\citeproc{ref-openid:vp}{15}{]}. This technique drastically
improves the control of principals in scenarios in which credentials
contain more claims than strictly required. It would be a strong
advantage of OpenID's new design, were it not that their architecture is
also the main \emph{cause} of such scenarios in the first place. By
splitting the wallet from the issuer -- issuing `reusable' credentials
from which presentations for multiple verifiers can be created -- it
indeed becomes necessary for the wallet to filter which claims are
disclosed to which verifiers. Other models, like OIDC, do not have this
issue to begin with: the issuer creates a new credential (presentation)
for each request, tailored to the verifier.

Neither informed consent, nor selective disclosure therefore add a real
advantage to OpenID's approach to decentralization; especially not when
compared to (other) federated models. With respect to control, this lack
of technological progress is not core issue, though, since the
regulations themselves do not allow for more (cf.
Section~\ref{sec-recentral}).

\subsection{Privacy}\label{sec-privacy}

The increase in \emph{privacy} claimed by OpenID is supposedly due to
the principal's ability to ``\emph{directly} present identity
information to the relying parties,'' who can ``receive and validate
presented credentials without {[}either the principal or the verifier{]}
directly interacting with the issuer''
{[}\citeproc{ref-openid:trustmodel}{61}{]}. In their white paper, they
say this `most notable feature' mimics physical credentials\footnote{OpenID's
  analogy -- that their trust model mimics physical credentials -- does
  not hold well under further scrutiny. While, at first sight, a
  physical credential (e.g., a driver's license) might seem to contain
  everything a verifier (e.g., a police officer) needs to know, this is
  not the case. It does not merely lists assertions (e.g., name, allowed
  type of vehicles), but -- similar to a digital credential -- it also
  contains different kinds of \textbf{attestation information}. It
  provides information about the \emph{authentication context} (e.g.,
  registration number, dates of issuance and expiry), as well as
  verification material for \emph{holder binding} (e.g., a picture,
  indication of sex and age) and \emph{issuer binding} (e.g., a
  signature, holographic mark, and other anti-fraud features). In
  practice, it is true that verifiers often merely compare the holder
  binding with the person in front of them, perhaps glance at some
  quick-to-check context details (e.g., expiry date), and rely on a
  general feeling of what such a document should look like. To warrant
  any real assurance in the claims, however, each of these factors needs
  to be thoroughly verified. In particular, all aspects of the issuer
  binding need to be checked, which requires up to date knowledge about
  the details involved; and the same goes for the authentication
  context. Especially the latter will, at least in case of licenses,
  typically include an interaction between verifier and issuer (e.g.,
  whether the license's registration number actually exists, and has not
  recently been revoked by a judge).}, since ``{[}issuers{]} no longer
know what activity {[}principals{]} are performing at which relying
party'' {[}\citeproc{ref-openid:trustmodel}{61}{]}. In particular, they
claim that ``scenarios where the {[}issuer{]} has no legitimate reason
to know which {[}relying party{]} the user wants to access resources
from and when they do so'' are not achievable with (other) federated
flows {[}\citeproc{ref-openid:trustmodel}{61}{]}.\footnote{Note that
  this concern also holds in the other direction: information about the
  issuer, included in a credential, might reveal information about the
  subject or holder to the verifier. This is especially the case when an
  issuer provides only one type of credentials, and/or is the only
  issuer providing such credentials. Privacy in this direction is much
  harder to achieve. The implicitly revealed knowledge might partially
  be reduced by grouping multiple issuers together -- possibly even
  using shared key materials {[}\citeproc{ref-openid:vci}{14}{]}.
  However, since the issuer is the root of trust for the verifier, this
  mitigation might lead to dilution of trust in the ecosystem
  {[}\citeproc{ref-openid:security}{46}{]}.}

Again, these are strong statements. The principal can indeed present
credentials directly to the relying party; but this is technically not
different from the OIDC authorization code flow: the ID token passes
from the issuer, through the user agent -- in this case a browser
instead of a wallet -- to the verifier. Importantly, however, in both
architectures, both the principal and the verifier have to interact with
the issuer.

First, the principal must at some point \textbf{retrieve the credential}
from the issuer, to store it in their wallet (or browser memory). In
theory, this could indeed happen long before the participant presents
the information to the relying party. However, OpenID's specifications
stress that for \emph{privacy considerations}, wallets ``should
\textbf{\emph{not}} store credentials longer than needed''
{[}\citeproc{ref-openid:vci}{14}{]}. In fact, since ``presentation
sessions \ldots{} can be linked on the basis of unique values encoded in
the credential,'' wallets are advised to use ``a \textbf{unique}
credential per presentation or per verifier''
{[}\citeproc{ref-openid:vci}{14}{]} to avoid such correlation -- ``each
with unique issuer signatures {[}and{]} keys'' -- and then discarding
the credential {[}\citeproc{ref-openid:vp}{15}{]}.\footnote{These
  considerations also negate OpenID's claims that their specifications
  make technical implementation easier, simpler, and more scalable, or
  that they would lead to a more seamless user experience
  {[}\citeproc{ref-openid:trustmodel}{61}{]}. Likewise, they make it
  hard to get any benefit from on-device storage with hardware-backed
  encryption -- as advocated for in their draft specification on
  security and trust in OpenID4VCI ecosystems
  {[}\citeproc{ref-openid:security}{46}{]}.}

Second, in order to \textbf{verify the issuer's signature}, the verifier
will have to contact the issuer \emph{upon receiving each presentation}.
While it is possible to cache some of the issuer's verification material
(e.g., public keys), this is equally true of (other) federated systems
like OIDC. Moreover, caching becomes useless in case the issuer and
wallet follow the advise to use a \emph{unique key} per credentials and
a \emph{unique credential} per presentation (cf.~supra).

Finally, even if OpenID's architecture would successfully prevent
issuers from learning about the user's activity at relying parties, it
would have achieved this by introducing a new intermediary entity -- the
wallet -- who will possess the same information instead. Whether this is
desirable or not will depend on the context. While the design of
OpenID4VCI and OpenID4VP can somewhat reduce the frequency of direct,
synchronous interactions, it is therefore fair to say that OpenID's
strong privacy claims are at least an exaggeration. To the contrary, the
considerations regarding the risk of correlation -- emphasized in the
specifications themselves -- make it more plausible that the proposed
flows are in fact not desirable for privacy after all.

\begin{center}\rule{0.5\linewidth}{0.5pt}\end{center}

In this section we critically assessed each of the benefits proclaimed
by OpenID to constitute a user-centric paradigm shift in digital
identity. We conclude that their claims about control, privacy, and
portability are either plainly incorrect, exaggerated, or already
present in (other) federated solutions. Moreover, we pointed out that
some of their so-called `strong points' in fact hardly make up for
certain less desirable effects of the new architecture.

That OpenID's claims do not survive adequate scrutiny also calls into
question whether their specifications are truly a good choice for
implementing the EUDI regulations. As such, there is nothing wrong with
a specification that is narrowly tailored to handful of specific use
cases: even with all its limitations and vulnerabilities taken into
account, OpenID's architecture still supports most traditional scenarios
of credential exchange. However, as a foothold for the EUDI
infrastructure, their design drastically limits the potential of the
EU's strategy, both in its limited capabilities and in its lack of --
backwards compatible -- options for evolution.

\section{The politics of control}\label{sec-politics}

Having established OpenID's inability to live up to its
paradigm-shifting promises, we take a look at a number of choices made
by the EU's regulations themselves. From their promotion materials, to
the regulations' recitals, it is clear that the EU's vision is also full
of big promises: ``Union citizens and residents in the Union should have
the right to a digital identity that is under their \emph{sole
control}'' {[}\citeproc{ref-eu:eudi}{7, p. art.3}{]}. Wallets aim to
give those citizens ``\emph{full control} on what data they share to
identify themselves with online services \ldots{} at all times''
{[}\citeproc{ref-eu:qa}{68}{]}. Users will also be able to \textbf{share
digital documents},\footnote{It remains to be seen whether the
  terminology used in the regulations is actually compatible with the
  exchange of documents that are not credentials of attributes about the
  wallet user -- who is both holder and subject. We reiterate that the
  choice for the OpenID architecture, as implementation of the EUDI
  framework, in any case fails to deliver this promise, since the design
  of OpenID4VCI makes it technically impossible (cf.
  Section~\ref{sec-vci}).} and ``\textbf{prove statements} {[}i.e.,
specific personal attributes{]} about themselves and their relationships
\textbf{with anonymity} (i.e.~\emph{without revealing identifying
data})'' {[}\citeproc{ref-eu:qa}{68}{]}. In other words, the EUDI
regulation promises far-reaching \textbf{self-sovereign identity} (SSI):
a digital identity model in which each individual controls who they are
on the Web, i.e., what information they are associated with when
interacting with online services.\footnote{While the main text of the
  EUDI regulation does not explicitly mention SSI, the EU's aims are
  clear in many of its preliminary documents. Following the introduction
  of the European Blockchain Services Infrastructure (EBSI), an
  evaluation report of eIDAS highlighted the latter's incompatibility
  with the former. Not only did it bring to the attention that ``some
  stakeholders consider identity data too sensitive to store centrally
  and suggest to consider decentralised systems for issuing trusted
  certificates based on distributed ledger technologies and
  self-sovereign identity solutions (SSI)''
  {[}\citeproc{ref-eu:evaluation}{69}{]}; it also refered to a survey,
  which indicated that ``a large majority of respondents (63\%) thought
  that it would be useful to own a secure single digital identity to
  access both public and private services and get control over the use
  of their data'' {[}\citeproc{ref-eu:evaluation}{69}{]}. Importantly,
  the evaluation concluded that ``the eIDAS Regulation does not enable
  this type of use cases \ldots{} {[}yet{]} an adaptation of the
  {[}section{]} on trust services to support the development of
  verifiable claims could'' {[}\citeproc{ref-eu:evaluation}{69}{]}. This
  evaluation was followed by several impact assessments
  {[}\citeproc{ref-eu:feedback1}{70}; \citeproc{ref-eu:feedback2}{71};
  \citeproc{ref-eu:impact}{72}; \citeproc{ref-eu:summary}{73}{]}, and an
  ENISA report exploring ``the potential of self-sovereign identity
  (SSI) technologies to ensure secure electronic identification and
  authentication to access cross-border online services''
  {[}\citeproc{ref-enisa:ssi}{74}{]}. Based on a number of research and
  pilot projects {[}\citeproc{ref-eu:bridge}{75};
  \citeproc{ref-essif:project}{76}{]}, all of them concluded positively
  that ``digital wallets are a practical way of implementing SSI''
  {[}\citeproc{ref-eu:impact}{72}{]}, which gives the user ``greater
  control over how its identity is represented to parties relying on the
  identity information and, in particular greater control over the
  personal information that it reveals to other parties''
  {[}\citeproc{ref-enisa:ssi}{74}{]}. The result was the initial
  proposal for what would become the EUDI regulation: ``Users expect a
  self-determined environment where a variety of different credentials
  and attributes can be carried and shared \ldots{} so-called
  self-sovereign app-based wallets managed through the mobile device of
  the user \ldots{} allowing for a secure and easy access to different
  services, both public and private, under his or her full control''
  {[}\citeproc{ref-eu:proposal}{77}{]}.}

Note that the SSI model does not imply \textbf{ownership} of the
identity information. The data itself can originate with another party,
and made available to the individual; the latter thus does not control
the availability of the information. This is echoed in the OpenID trust
model, which emphasizes that ``it is still up to the verifier to decide
whether to accept those credentials {[}and{]} it is still up to the
issuer to decide whether to issue the credential to the {[}individual{]}
in the first place {[}or{]} to revoke and invalidate the credential''
{[}\citeproc{ref-openid:trustmodel}{61}{]}. The sovereignty of the
individual therefore lies in their power to autonomously control who can
access which of the \emph{available} information under which
circumstances. Even with an ideal technological implementation, however,
it remains to be seen whether the legal requirements, formulated in the
EUDI regulation, are actually compatible with such a form of
self-sovereignty. In the following sections, we will look into the
concept of trusted lists, and why they are both a necessity and a risk
for internet freedom and security.

\subsection{Trusted lists}\label{sec-lists}

Proposed by the European Telecommunications Standards Institute (ETSI)
{[}\citeproc{ref-etsi:trustlists}{78}{]}, \textbf{trusted lists} are
perhaps the most pervasive regulatory device introduced in eIDAS, and --
in particular -- enforced in EUDI. These lists, published and maintained
by \emph{Trusted List Providers} (TLPs), contain the trust anchors
(i.e., identifiers and public keys) of regulated \emph{Trust Service
Providers} (TSPs). The \emph{trust services}, which these entities
provide, include a variety of \emph{digital identity solutions}
{[}\citeproc{ref-oecd:govdi}{28}{]}: (electronic) services involved in
authentication processes; e.g., the issuance, validation, and
verification of signed attestations (certificates), as well as their
life cycle management, and that of their verification material (e.g.,
timestamps, ledgers, signatures, and seals)
{[}\citeproc{ref-eu:eidas}{6}; \citeproc{ref-eu:arf}{8}{]}.\footnote{Without
  diving to deep into the nature of \textbf{trust} -- which is also
  undefined in the regulations -- and given the techno-economic context,
  we take it to be \emph{a relationship between two parties, that
  supports the exchange of data and services}
  {[}\citeproc{ref-idlab:trust}{79}{]}. In Section~\ref{sec-politics},
  we look into this in more detail.}

In particular, the European Commission maintains trusted lists of
different types of TSPs -- including wallets, attestation providers,
signature services, and other certification authorities -- that are
granted the \textbf{qualified} status,\footnote{While there is no
  trusted list for relying parties -- because their expected number
  would make this ``practically infeasible'' -- they are subjected to a
  similar system, in which they ``{[}receive{]} an access certificate
  from an \emph{access certificate authority}, {[}which{]} allows a
  wallet unit to authenticate {[}them{]}''
  {[}\citeproc{ref-eu:arf}{8}{]}.} and are thereby approved to provide
\textbf{qualified trust services}
{[}\citeproc{ref-eu:arf}{8}{]}.\footnote{To be precise, the eIDAS and
  EUDI regulations distinguish different types of attestations, each of
  which is assigned a \textbf{level of assurance} (LOA), determined by
  the digital identity solutions employed by their issuer, indicating
  the extent to which the relying party can be confident in the claimed
  information {[}\citeproc{ref-eu:eudi}{7};
  \citeproc{ref-oecd:govdi}{28}; \citeproc{ref-eu:assurance}{80}{]}.
  Non-qualified EAAs can only be issued with LOA `low' to `substantial',
  while qualified EAAs (\textbf{QEAAs}) are the only ones that can have
  LOA `high'. The legal significance of this difference is big: in
  contrast to EAAs, QEAAs are \emph{legally equivalent} to documents
  with a handwritten signature. Two categories of QEAAs are treated
  separately: \emph{official documents}, issued by a \emph{public sector
  body} responsible for an \emph{authentic source} (\textbf{PuB-EAAs});
  and \emph{Person Identification Data} (\textbf{PID}), i.e., official
  \emph{identity documents} that ``{[}enable{]} the establishment of the
  identity of a natural or legal person''
  {[}\citeproc{ref-eu:attestations}{81}{]}.} To be put on a these lists,
TSPs need to adhere to certain requirements
{[}\citeproc{ref-eu:eudi}{7}, Annex V{]}, and be registered by a
\emph{registrar}: a supervisory body of their Member State, which in
turn notifies the European Commission
{[}\citeproc{ref-eu:notifications}{82};
\citeproc{ref-eu:certifications}{83}{]}. In the case of attestation
providers, this registration has to be renewed on every change in the
issued credentials, since the registered data includes ``the attestation
type(s) that the provider intends to issue to wallet units''
{[}\citeproc{ref-eu:arf}{8}{]}.

Surprisingly, this pervasive aspect of the regulation did not meet much
general opposition. One specific use of trusted lists, however, caused
an intense debate between the legislature and the internet community --
in particular with browser vendors and advocates for net-neutrality. The
topic under discussion was the \textbf{qualified website authentication
certificate} (QWAC): an electronic attestation, in the form of a TLS
certificate\footnote{A Transport Layer Security (TLS) certificate
  encrypts one entire internet connection, on a lower architectural
  layer than the application layer, thereby protecting the integrity of
  the data transmitted between the client and server applications using
  that connection. Since TLS certificates are linked to domain names,
  they also prove the origin of the data, and are therefore used to
  authenticate the domain owners to the user.} -- or an attribute
certificate cryptographically linked to one (ac-QWAC)
{[}\citeproc{ref-ietf:rfc5755}{84}{]} -- and issued to EU-based, audited
TSPs that comply with a number of ``minimal security and liability
obligations'' {[}\citeproc{ref-eu:eudi}{7}{]}.

Supported by a report from the European Union Agency for Cybersecurity
(ENISA\footnote{Previously known as the European Network and Information
  Security Agency.}), on critical improvements for website
authentication {[}\citeproc{ref-enisa:qwacs}{85}{]}, QWACs were
introduced in eIDAS as a voluntary choice for websites. Their intent was
to ``provide a means by which a visitor to a website can be assured that
there is a genuine and legitimate entity standing behind the website,''
and as such to ``contribute to the building of trust and confidence in
conducting business online'' {[}\citeproc{ref-eu:eidas}{6}{]}. The
regulation only became an issue after a crucial amendment in EUDI
{[}\citeproc{ref-eu:eudi}{7}, art. 45{]}, requiring browsers to treat
QWAC providers as root CAs -- and thus recognize and display these
government-approved certificates to their users as an assurance of
trustworthy services -- \emph{regardless of the well-established
security measures traditionally upheld by the browsers themselves}
{[}\citeproc{ref-cepis:eudi}{86}{]}. In the following sections, we
explore the two main consequences of this decision, that caused the
worldwide criticism against it.\footnote{In response to the criticism,
  the following paragraph was added to the final EUDI regulation: ``In
  order to contribute to the online security of end-users, providers of
  web-browsers should, in exceptional circumstances, be able to take
  precautionary measures that are both necessary and proportionate in
  reaction to substantiated concerns regarding security breaches or the
  loss of integrity of an identified certificate or set of
  certificates'' {[}\citeproc{ref-eu:eudi}{7}{]}. This amendment should
  ensure that the regulation ``{[}does not{]} affect the freedom of
  web-browser providers to ensure web security \ldots{} in the manner
  and with the technology they consider most appropriate''
  {[}\citeproc{ref-eu:eudi}{7}{]}. However, the paragraph only appears
  in the preliminary recitals of EUDI, and is thus -- unlike the article
  itself -- not legally binding {[}\citeproc{ref-cepis:eudi}{86}{]}.}

\subsection{A weakened internet security}\label{sec-weakened}

TLS certificates can validate several characteristics of an internet
connection, including the domain name and organization linked to the
website. Like with any certificate, trust in TLS verification is
dependent on trust in the CA vouching for it. Since users can hardly be
expected to know which of the many CA's to trust on the Web, browser
vendors take up this burden by maintaining a `trusted list' of vetted
CAs, typically based on advice of the Certification Authorities and
Browsers (CA/B) Forum {[}\citeproc{ref-isoc:impact}{87}{]}.\footnote{These
  lists of trusted CAs, also called \emph{root stores}, are carefully
  compiled based on well-defined, transparent \textbf{root programs}:
  security standards, including policies and safeguards concerning
  encryption, public transparency, audits, and other practice
  requirements {[}\citeproc{ref-mozilla:position}{88}{]}.} In
particular, until the early 2020s, websites complying with a strict set
of Extended Validation (EV) rules -- compiled by this CA/B Forum -- were
displayed with a `green shield' indicator in the address bar of many
browsers, to indicate this to the user.

The eIDAS certificates, and EUDI's obligation to display them in
browsers, copy this design, but rely on government certification instead
of a multi-stakeholder CA architecture. However, while EUDI literally
states that ``the results of existing industry-led initiatives, for
example the {[}CA/B{]} Forum, have been taken into account''
{[}\citeproc{ref-eu:eidas}{6}{]}, the CA/B community -- including most
major browsers -- had in fact already deprecated or discontinued the EV
system by the time EUDI was drafted. They moved away from this practice
after research by Google and UC Berkeley had shown that the indicators
did not have a significant effect on the behavior of users: they
``{[}did not{]} provide users with clear, actionable cues about online
trust, and were therefore adding cost and complexity but little or no
benefit'' {[}\citeproc{ref-isoc:impact}{87}{]}.

The cost--benefit analysis was, however, not the reason for the strong
reaction which the regulation triggered. Responding to the proposal's
feedback rounds {[}\citeproc{ref-eu:feedback1}{70};
\citeproc{ref-eu:feedback2}{71}{]}, the Common Certification Authority
Database (CCADB) {[}\citeproc{ref-ccadb:qwacs}{89}{]} -- an initiative
of the Linux Foundation -- together with major browsers, argued that
mandating QWACs undermines technical neutrality, interoperability, and
user privacy -- principles that are central to the intent of eIDAS
itself. Similar objections were later raised in an impact brief from the
Internet Society (ISOC), which claimed that ``ETSI's assumption that
browsers would add eIDAS-approved {[}TSPs{]} to the trusted root list
violates {[}the ideal of an open, trustworthy Internet{]},'' based on
principles of collaboration, expertise, transparency, and consensus
around ``trust criteria mutually agreed among \ldots{} relevant
stakeholders'' {[}\citeproc{ref-isoc:impact}{87}{]}.

In a position paper {[}\citeproc{ref-mozilla:position}{88}{]},
reiterating the objections of their earlier response
{[}\citeproc{ref-mozilla:response}{90}{]}, Mozilla called the new
regulation an `unprecedented move', which ``will amount to forced
website certificate whitelisting by government dictate and will
irremediably harm users' safety and security''
{[}\citeproc{ref-mozilla:position}{88};
\citeproc{ref-mozilla:response}{90}{]}. They emphasized that the
proposal ``goes against established best practices \ldots{} created by
consensus from the varied experiences of the Internet's explosive growth
{[}which{]} have successfully helped keep the Web secure for the past
two decades'' {[}\citeproc{ref-mozilla:response}{90}{]}. The obligation
for browser vendors to automatically include TSPs in their root
programs, would effectively ``replace the security expertise of major
browser companies \ldots{} with legislation premised on weaker and
discredited security architectures,'' leading to ``a regression in the
security assurances that users have come to expect from their browsers''
{[}\citeproc{ref-mozilla:position}{88}{]}. The requirements for
inclusion in, for example, Mozilla's root program, are more rigorous,
and more transparent, providing for more public oversight, and more
stringent audits than eIDAS' criteria for TSPs
{[}\citeproc{ref-mozilla:position}{88}{]}. Therefore, ``by mandating
that TSPs be supported by browsers in general, and in particular when
they fail to meet the security and audit criteria of {[}the browsers'{]}
root program, {[}EUDI{]} will negatively transform the website security
ecosystem in a fundamental way''
{[}\citeproc{ref-mozilla:position}{88}{]}.

According to ISOC's impact brief, these negative effects come in two
forms: by issuing incorrect certificates, and through inability to
rapidly address security incidents {[}\citeproc{ref-isoc:impact}{87}{]}.
This risk assessment is echoed in an open letter published by the
Electronic Frontier Foundation (EFF), signed by numerous cybersecurity
researchers, advocates, and practitioners: ``allowing some website
certificates to bypass existing security standards \ldots{} increases
the risk that insecure or malicious certificates will be issued \ldots{}
and {[}makes{]} it impossible for the cybersecurity community to quickly
respond when certificates are found to pose a risk''
{[}\citeproc{ref-eff:letter}{91}{]}. In general, the EFF characterizes
the regulation as ``a dangerous cybersecurity policy trend,'' which goes
against established norms in cybersecurity and risk management, and
``compels private actors to forgo their duty to those who use their
products and services, by assuming that because government-appointed
{[}CAs{]} are subject to government security standards, they can pose no
cybersecurity risk'' {[}\citeproc{ref-eff:letter}{91}{]}.

\subsection{(Re)centralized (dis)trust}\label{sec-recentral}

The trusted lists concerning website authentication share a lot of
similarities with their counterparts for providers of wallets,
attestations, and signatures -- in fact, with any kind of eIDAS/EUDI
lists.\footnote{For example, the promotion of a central key management
  system (CKM), through lists of trusted signature services, would have
  the majority of citizens entrust cryptographic information (e.g.,
  private keys) to government-selected organizations. Without careful,
  open, industry-led security measures, this will undermine citizen's
  autonomy and control, rather than bolster it. Consequences of such
  approaches to ``let there be trust'' by government decree can, for
  example, be found in the aftermath of similar legislation in Uruguay
  {[}\citeproc{ref-sabiguero:trust}{92}{]}.} In general, each of these
trusted lists institutionalizes an \emph{elevated status}, granted
\emph{by national government registrars} to a select group of TSPs
{[}\citeproc{ref-eu:certifications}{83}{]}. This entails both economical
and political risks.

The EUDI design allows only registered systems -- abiding by the
criteria set forth in the regulations -- to participate in the EUDI
ecosystem, and interact with other parties via the architecture's
protocols {[}\citeproc{ref-eu:protocols}{93}{]}. Therefore, ``those who
are on the list receive economic advantages; those who are not on it
have disadvantages'' {[}\citeproc{ref-schall:eudi}{94}{]}. In itself,
this is not an issue. However, several of the requirements can form
\emph{economical obstacles} (e.g., increased liability, financial
minima) or \emph{technical hurdles} (e.g., renewal for each changed
attestation type) {[}\citeproc{ref-eu:certifications}{83};
\citeproc{ref-eu:protocols}{93}{]}. This could lead to an ecosystem that
\emph{favors affluent organizations}, strengthening their
\emph{privileged position} in their respective techno-economic spheres,
which increases risks like vendor lock-ins and gate-keeper behavior
{[}\citeproc{ref-schall:eudi}{94}{]}. Rather than a decentralization of
digital identity, such an ecosystem could therefore result in a
\textbf{(re)centralization} instead, in which authentication on the Web
is directed by a small number of key economical and political actors.

Importantly, this not only holds for the TSPs themselves, but \emph{also
for relying parties} making use of their services
{[}\citeproc{ref-eu:relyingparties}{95}{]}. To obtain even the simplest
attestation from a wallet, a relying party would need to jump through
the hoops described above, \emph{regardless of the choice of the wallet
user}. This conclusion definitely falls short of the promise for a
`user-oriented identity management' that would put users `in charge' of
their own personal information. Indeed, while ``people can choose which
bits of their identity information to share''
{[}\citeproc{ref-igrant:eudi}{57}{]}, they have \emph{nothing to say}
about which issuer can provide their credentials, nor to which verifiers
they can present their attributes. As such, wallets merely provide a
uniform interface through which only registered parties -- willing to
pay the cost of entry -- can exchange information.

Moreover, many of the implementing regulations require TSPs to \emph{log
and preserve information} about their systems and the parties they
interact with {[}\citeproc{ref-eu:information}{96};
\citeproc{ref-eu:breaches}{97}; \citeproc{ref-eu:matching}{98}{]}. In
certain cases (e.g., security breaches), TSPs have the obligation to
\emph{notify the European Commission} -- upon which the latter can
decide to \emph{suspend the service/provider} from the ecosystem. These
requirements are not limited to information about TSPs, wallets, and
relying parties: in some cases -- e.g., \emph{identity matching} by
cross-border services -- \emph{personal information} about the wallet
user, or subject of authentication, must also be logged.

Given the EUDI's possible tendency towards (re)centralization
(cf.~supra), these increased logging requirements could pose an
additional risk of abuse; e.g., \emph{monitoring the usage of wallets}
to \emph{profile citizens' behavior} and interests -- either by
malicious TSPs themselves, or as a target of external actors. Such risks
not only negate OpenID's emphasis on minimizing unnecessary data
disclosure {[}\citeproc{ref-openid:vci}{14};
\citeproc{ref-openid:vp}{15}; \citeproc{ref-igrant:eudi}{57}{]}; they
also stand in \emph{sharp contrast with other EU legislation}, including
data protection regulations like the GDPR, and the EUDI regulation's own
insistence on \textbf{unlinkability}. The latter is particularly
surprising, since it is part of the same regulation, and specifically
aims to prevent \emph{any} party to bring together ``data that allows
for tracking, linking, correlating or otherwise obtain knowledge of
transactions or user behavior \emph{unless explicitly authorized by the
user}'' {[}\citeproc{ref-eu:eudi}{7}{]}.\footnote{Similar to the case of
  QWACs, the regulators' response was limited to a minor addition to the
  (not legally binding) recitals of the final EUDI regulation
  {[}\citeproc{ref-eu:eudi}{7}{]}. While this once more emphasizes the
  need for \emph{unobservability} of a user's transactions by wallet
  providers, it in no way answers the question of its technical
  implementation, given the practically opposite requirements
  {[}\citeproc{ref-cepis:eudi}{86}{]}.} This is precisely the tension
pointed out by the expert group for Legal and Security Issues (LSI) of
the Council of European Professional Informatics Societies (CEPIS):
``Unfortunately the development of the Architecture Reference Framework
(ARF) \ldots{} is not transparent, but behind closed doors, and
\textbf{\emph{available drafts do not support the legal requirements for
safeguards for unlinkability}}'' {[}\citeproc{ref-cepis:eudi}{86}{]}.

Most importantly, however, the economical and political design of EUDI,
as described above, allows governments to whitelist favored service
providers, without much restriction. CEPIS criticizes that there are
``no safeguards that prevent the governments \ldots{} from exercising
\emph{surveillance over everything its users do with it}''
{[}\citeproc{ref-cepis:eudi}{86}{]}. It is important to realize that
these scenarios are not `horror stories', but rather ``deductions from
already known types of attack, economic incentives and \ldots{}
historical experience'' {[}\citeproc{ref-schall:eudi}{94}{]}.
Organizations managing CA root programs, like Mozilla, have first hand
experience with these dangers. They list, amongst others, the
governments of Mauritius, Kazakhstan, and China, as ``authoritarian
regimes {[}who{]} have long sought to override {[}trusted list{]}
policies'' {[}\citeproc{ref-mozilla:position}{88}{]}.

Note that none of the critics referred to above are against the concept
of trusted lists. In fact, they are an \textbf{architectural necessity}
in any decentralized system. At the same time, however, they are an
instrument of power, which should urge us to ask questions like: ``In
whose hands does this tool fall?'', ``How strong are the barriers
against misappropriation, commercialization, and criminal
exploitation?'', ``How is abuse prevented if political or economic
interests are involved?'' {[}\citeproc{ref-schall:eudi}{94}{]}. After
all, ``a globally connected Internet is premised on the ability of
Internet users to access and use resources in other networks
\emph{without unnecessary restrictions}''
{[}\citeproc{ref-isoc:impact}{87}{]}. Giving political actors the power
to directly influence such restrictions -- either by imposing them or by
overruling them -- sets a dangerous antecedent.

\section{Alternative solutions}\label{sec-alt}

In this final section, we highlight some of the alternative solutions
already mentioned earlier, and propose a number of additional lines of
research worth looking into. Since immediate legislative changes are
improbable, we focus on technological specifications that might overcome
some of the issues we raised, and pave the way towards a truly
self-sovereign digital authentication framework.

To achieve this, we argue that solutions should be aimed at a broad,
explicitly defined interpretation of authentication and identity, as put
forward in Section~\ref{sec-char} -- or should at least provide
sufficient points of extensibility to enable probable evolution paths
towards it. The core principles of the semantic Web are a good starting
point: global identifiers (i.e., URIs, and in particular DIDs
{[}\citeproc{ref-w3c:did}{10}{]}), combined with semantically rich
structures like RDF {[}\citeproc{ref-w3c:rdf}{100}{]}, allow for a
far-reaching interoperability with existing and future technologies.

Based on this strong foundation, \textbf{W3C's models} of
\emph{verifiable credentials and presentations}
{[}\citeproc{ref-w3c:vc}{11}{]} -- also supported by OpenID -- offer the
most potential for aligning the wide variety of electronic attestations
of attributes. As we pointed out in Section~\ref{sec-trust}, they are
not inherently more expressive or secure than classic OIDC tokens, but
they achieve those features in a semantically richer, more
interoperable, and extensible manner. On the other hand, while W3C VCs
\emph{can handle partial identity} (e.g., roles, pseudonyms
{[}\citeproc{ref-idlab:pseudo}{101}{]}), they are still targeted to a
subject, and can therefore \emph{cannot express certificates} in the
broader sense. A compatible model for such general `attestation
documents' therefore remains a crucial topic for future research.

Throughout Section~\ref{sec-arch}, we already mentioned several other
OpenID specifications that together are functionally equivalent to
OpenID4VCI and OpenID4VP. Based on the well-established OIDC
{[}\citeproc{ref-openid:connect}{3}{]}, SIOPv2 aligns credentials with
self-issued presentations {[}\citeproc{ref-openid:siop}{60}{]}, OIDC4IDA
adds a levels of assurance model {[}\citeproc{ref-openid:ida}{102};
\citeproc{ref-openid:idaschema}{103};
\citeproc{ref-openid:idareg}{104}{]}, and OIDC Claims Aggregation can
substitute for wallet-like behavior
{[}\citeproc{ref-openid:aggregation}{67}{]}. Contrary to OpenID4VCI and
OpenID4VP, the latter specifications rely on existing standards where
possible, and provide strong extensibility and interoperability -- both
between themselves and with external specifications.

However, we also pointed out several problems with the OpenID ecosystem
in general, which should be addressed in any possible alternative. In
particular, attention needs to be payed to best practices in internet
security and software design -- such as the separation of orthogonal
concerns. Ideally, an alternative solution would therefore be based
directly on OAuth 2.1 {[}\citeproc{ref-ietf:oauth21}{52}{]}. This would
lift OpenID's restriction to a single endpoint, and instead allow the
immense variety of attestations to be exchanged via any kind of
interface: classic RESTful APIs, database queries, web streams etc.

To access such heterogeneous forms of attestations, however, another
issue needs to be addressed first. In Section~\ref{sec-vci} and
Section~\ref{sec-vp}, we criticized the inadequacy of a single string to
express the entire semantics of each type of credential -- even for the
limited variety exchangeable through OpenID wallets. We also pointed out
the issues with OpenID's several custom, non-interoperable query
languages (i.e., DCQL and claim descriptions). Within the restrictions
of OpenID, the JSONPath or JSON Schema specifications
{[}\citeproc{ref-ietf:rfc9535}{53};
\citeproc{ref-ietf:jsonschema}{105}{]} offer a more standardized,
interoperable alternative. However, as we have pointed out in
Section~\ref{sec-oidc}, even a standardized approach based on JSON has a
limited flexibility and interoperability in the light of a more
heterogeneous range of credentials. The suggested addition of SPARQL
queries {[}\citeproc{ref-w3c:sparql}{106}{]}, to request specific claims
from the combined credentials in a wallet, would already make a big
difference {[}\citeproc{ref-idlab:queryvc}{54}{]}. In order to also
include other data interfaces, research into a more abstract query
(meta-)language is necessary.

A choice for OAuth would also open up the possibility of using its
\emph{User-Managed Access} (UMA) extension
{[}\citeproc{ref-uma:grant}{107}; \citeproc{ref-uma:fed}{108}{]}. By
modeling a dynamic negotiation with the verifier, UMA enables more
complex authorization contexts to be established, thereby lifting
OpenID4VCI's limitation to static, preset credential configurations (cf.
Section~\ref{sec-vci}). UMA also emphasizes asynchronous interaction
with the principal, which opens up the way for use cases involving
automation. In earlier work, we already discussed UMA in more detail as
an alternative to approaches involving OIDC and access control lists
{[}\citeproc{ref-idlab:uma}{109}{]}. We also provided a profile
specification for UMA, called \emph{Authorization for Data Spaces}
(A4DS) {[}\citeproc{ref-idlab:a4ds}{110}{]}, which puts forward an
authorization model in which wallets let users regulate access to data
that remains safely at its original source -- functioning more like
remote keys than like portable hard drives.

Based on the pointers in this section, one other interesting project is
the \emph{Grant Negotiation and Authorization Protocol} (GNAP)
{[}\citeproc{ref-ietf:rfc9635}{111}{]}. This draft specification --
sometimes hailed as `OAuth 3.0' -- combines two decades of best
practices around Oauth 2.0 with an OIDC-like identity provider and an
authorization model inspired by UMA. It is important to keep in mind,
though, that neither GNAP, nor any other alternative, will be able to
actually fulfill the promise of a decentralized, self-sovereign European
identity framework, as long as the risks in Section~\ref{sec-politics}
-- caused by the economic and political impact of trusted lists (cf.
Section~\ref{sec-lists}) -- are not addressed. Only through transparent
collaboration with experts and stakeholders can the technological
potential of these solutions be realized.

\section*{Summary and conclusion}\label{summary-and-conclusion}
\addcontentsline{toc}{section}{Summary and conclusion}

We started from the observation that, in legal and technical sources
pertaining to identity in data spaces -- in particular those related to
recent EU regulations -- the term `\textbf{identity}' is ill-defined.
Aggregating several older international standards' glossaries, in
Section~\ref{sec-char} we settled on a workable definition: \emph{an
identity is a subset of an entity's characteristics, sufficient to
represent and recognize that entity}. We also clarify the concept
`\textbf{authentication}', which is the \emph{process of determining the
authenticity of information, by verifying its origin and integrity},
often through a (cryptographic) binding between multiple claims. In
Section~\ref{sec-trust}, we explained the roles of these concepts in the
trust relations of the \emph{issuer-holder-verifier} model. We also
showed how practically all verifiable information -- any certificate --
can be identity information (i.e., a credential); in particular in the
light of \emph{context-dependent} (partial) identities like roles,
pseudonyms, and anonymity.

Applying our findings to OpenID's architecture for the EUDI framework,
in Section~\ref{sec-arch} we exposed several issues with OIDC,
OpenID4VCI, and OpenID4VP, indicating a mismatch between a general
conceptualization of identity and the limited capabilities of those
specifications. Implementing the EUDI framework in an
\textbf{OIDC}-based architecture precludes a healthy separation of
concerns, leaves the door open for insecure practices, and immediately
limits the flexibility of the provided interfaces (cf.
Section~\ref{sec-oidc}). While \textbf{OpenID4VCI} and
\textbf{OpenID4VP} replace identity tokens with credentials, their
reliance on preset configurations, lack of expressivity in credential
types, limited query language, and format-dependent claim semantics
drastically narrow the use cases to which they can be applied.
Furthermore, since their credentials must have an (interactively)
identified subject, the possibilities for automated scenarios are
limited (cf. Section~\ref{sec-vci} and Section~\ref{sec-vci}).

In Section~\ref{sec-paradigm}, we discussed the strong \emph{parallel}
between OpenID4VP and OpenID4VCI, comparing them to self-issued
credentials in SIOPv2 (OIDC), and self-asserted claims in W3C's
Verifiable Presentations. We find that the latter specifications build
direct, holder-based issuance \emph{on top of} -- and compatible with --
their respective third-party issuance flows, while the former
specifications provides similar functionally through \emph{almost
identical yet incompatible mechanisms}. Looking for the rationale behind
this double tiered design, we discussed the \textbf{paradigm shift} --
claimed by OpenID -- away from federated models and towards
user-centricity. We refuted its claimed increase in user \emph{control,
privacy, and portability} of identity information; and concluded that
the OpenID trust model in no aspect goes beyond existing decentralized
approaches:

\begin{itemize}
\item
  While an increase in \emph{offline functionality} could be attributed
  to OpenID's new architecture, the BYOI \textbf{portability} it
  proclaims is not a real innovation, since this benefit is already
  present in many other decentralized systems -- including OpenID's own
  OIDC {[}\ref{sec-portability}{]}. The same goes for the facilitation
  of \emph{informed consent} to bolster people's control over their
  personal information.
\item
  \emph{Selective disclosure}, on the other hand -- providing
  fine-grained \textbf{control} over individual claims in larger
  credentials -- predominantly applies to \emph{scenarios that arise
  from OpenID's design itself}. In other decentralized models, this form
  of data minimization is much less of a necessity
  {[}\ref{sec-control}{]}.
\item
  The introduction of a user-bound wallet also does not provide the
  promised increase in \textbf{privacy}. OpenID provides no reason why a
  wallet provider would protect the private information in credentials
  better than their issuer. Moreover, this new intermediary does not
  truly make \emph{direct interaction} between issuers and verifiers
  redundant -- contrary to OpenID's claims. In fact, when taking into
  concern OpenID's own privacy considerations, the resulting flow is
  \emph{practically identical to OIDC} {[}\ref{sec-privacy}{]}.
\end{itemize}

From this analysis, we concluded that OpenID's specifications offer a
technological framework that is tailored to a specific set use cases --
including most \emph{classic scenarios of credential exchange} --
without necessarily offering more than other decentralized identity
models. Since this \emph{limits the capabilities and interoperability}
of the ecosystem, we called into question whether OpenID is truly a good
choice as foothold for EUDI.

In Section~\ref{sec-politics}, we looked into a number of legislative
choices of the European regulations themselves, and discussed their
impact on the EU's promise of \emph{self-sovereign identity}. In
Section~\ref{sec-lists}, we explained the introduction of trusted lists,
assurance levels, and qualified service providers; as well as their
increasingly strong constraints -- and ties to national governments --
through the EUDI's amendments to the eIDAS regulation. As a concrete
example, in Section~\ref{sec-weakened} we highlighted the commotion in
the internet community around \emph{qualified website authentication
certificates} -- called out by many as a dangerous trend in
cybersecurity policy, weakening global internet security by violating
established norms on net-neutrality and privacy. Extrapolating these
implications to trusted lists in general, we emphasized the risks of
making participation in an advantageous market dependent on a
registration procedure that forms an \emph{economical cost of entry},
and is \emph{governed by political institutions}. The economic and
political incentives in such ecosystems not only risk a de facto
\textbf{recentralization of digital identity}, driven by
\emph{gate-keeper behavior} of privileged market players; but, without
sufficient safeguards, they are also vulnerable to government abuse.

Towards individual people, the regulations fall equally short of their
goal. As we discussed in Section~\ref{sec-recentral}, the exclusory
ecosystem \emph{severely restricts people} in what they can do with
their personal information. Because the ecosystem is limited to
registered parties, they \emph{cannot freely choose} which issuers and
verifiers to interact with, even if the latter meet all technical
requirements. Such an ecosystem thus hardly fulfills the promise of
\emph{user-oriented identity management} in which individuals are
\emph{in charge}. Moreover, given the regulations' increased
requirements for \emph{logging and preserving information} -- and
disclosing it to government institutions -- this tendency towards
recentralization poses the additional risk of \emph{monitoring citizens'
behavior and interests}, regardless of the privacy-preserving
capabilities of the chosen technical framework. This stands in sharp
contrast with other legal requirements (e.g., GDPR), and promises about
the \emph{unlinkability} of correlating data.

We wrapped up this paper in Section~\ref{sec-alt} by arguing that the
risks related to EUDI will not diminish until the regulators revise the
problematic use of trusted lists. In anticipation of that, we gave a
number of suggestions for research into technical solutions that
overcome some of the issues with the architecture of OpenID. Leveraging
the extensibility and interoperability of semantic Web models (URIs,
DIDs, RDF), we take W3C VC's to be a good foundation, even though they
lack support for attestation documents that are not subject-based. We
repeated the functional equivalence between the new OpenID4VCI and
OpenID4VP specifications on the one hand, and a combination of the
already existing OIDC, SIOPv2, OIDC4IDA, and OIDC Claims Aggregation on
the other. However, we advocate for an architecturally sounder framework
based directly on OAuth, which has much broader applicability,
especially when extended with the asynchronous and dynamic features of
UMA, and their mutual evolution path through GNAP. Future research
should build on these, in line with the A4DS profile, and needs to look
into query (meta-)languages that are better suited to uniformly request
attestations from a more heterogeneous range of providers.

\begin{center}\rule{0.5\linewidth}{0.5pt}\end{center}

We conclude that, while OpenID's new architecture can indeed implement
the limited amount of use cases which the EUDI regulations allow, it
inherits an architectural history of ignoring best practices in software
design, internet security, and interoperability, without offering any
innovation: both the technical functionality (e.g., semantic
expressivity, query languages, automation) and the non-functional
features highlighted by the design (i.e., increase in control, privacy,
and portability) are already achievable with equivalent decentralized
solutions existing today, such as the OIDC extensions for self-issuance,
identity assurance, and claims aggregation.

Furthermore, without a thorough analysis of (digital) identity, the
regulations and their architecture have restricted themselves to a
narrow intuition of what credential exchange can look like -- thereby
missing the opportunity to construct a truly uniform model of
authentication on the Web. Alternatives, that aim at a broader
interpretation of authentication and identity, include the combination
of OAuth 2.1 with UMA, and GNAP. In either case, future research -- in
particular into uniform query (meta-)languages -- is needed to achieve
less trivial use cases.

However, even a more capable technical framework cannot surpass the
restrictions imposed by the EUDI regulations themselves. The manner in
which the regulators have institutionalized EUDI -- in particular
through the use of trusted lists -- risks to create perverse economic
and political incentives. Rather than building an inclusive
decentralized ecosystem of self-sovereign authentication on the Web,
these incentives can lead to a recentralization of digital identity, and
to dangerous vulnerabilities for abuse (e.g., profiling of citizens'
behavior). In its current state, the EUDI ecosystem therefore risks to
become a restriction of people's choice and control, rather than an
environment in which they feel in charge of their personal information.
Instead, authentication then becomes ``an entry ticket to a system in
which every transaction is clearly assigned to a person''
{[}\citeproc{ref-schall:eudi}{94}{]}.

\section*{References}\label{bibliography}
\addcontentsline{toc}{section}{References}

\protect\phantomsection\label{refs}

\end{document}